\documentclass[fleqn,usenatbib]{mnras}

\usepackage{newtxtext,newtxmath}

\usepackage[T1]{fontenc}

\DeclareRobustCommand{\VAN}[3]{#2}
\let\VANthebibliography\thebibliography
\def\thebibliography{\DeclareRobustCommand{\VAN}[3]{##3}\VANthebibliography}

\usepackage{multirow}

\usepackage{graphicx}	
\usepackage{amsmath}	
\usepackage[table,xcdraw]{xcolor}

\title[Morphology across cosmic time]{Morphology across cosmic time: assessing the evolution and interplay of disk and bulge-dominated galaxies in the CANDELS survey}

\author[Sampaio et al. 2023]{V. M. Sampaio,$^{1,2,3}$\thanks{E-mail: vitorms999@gmail.com}, I. Kolesnikov,$^{4}$, R. R. de Carvalho,$^{2,5}$, I. Ferreras$^{6,7}$, J. Silk$^{8,9,10}$
\\
$^{1}$ Instituto de Física, Universidad Técnica Federico Santa María, Av. España 1680, Valparaíso, Chile \\
$^{2}$ Núcleo de Astrofísica, Universidade Cidade de São Paulo, Rua Galvão Bueno 868, 01506-000, SP, Brazil\\
$^{3}$Millennium Nucleus for Galaxies (MINGAL), Chile \\
$^{4}$ Independent Researcher\\
$^{5}$ Instituto de Astronomia, Geofísica e Ciências Atmosféricas, Universidade de São Paulo, Rua do Matão 1226, 05508-090, São Paulo, SP, Brazil \\
$^{6}$ Instituto de Astrof\'isica de Canarias, Calle V\'ia L\'actea s/n,
E38205, La Laguna, Tenerife, Spain\\
$^{7}$ Department of Physics and Astronomy, University College London, Gower Street, London WC1E 6BT, UK \\
$^{8}$ William H. Miller III Department of Physics and Astronomy, Johns Hopkins University, Baltimore, Maryland 21218, USA \\
$^{9}$ Institut d’Astrophysique de Paris, UMR 7095 CNRS and UPMC, Sorbonne Universitè, F-75014 Paris, France \\
$^{10}$ Department of Physics, Beecroft Institute for Particle Astrophysics and Cosmology, University of Oxford, Oxford OX1 3RH, United
Kingdom
}

\date{Accepted XXXX Received XXXX; in original form XXXX}

\pubyear{2024}

\begin{document}
\label{firstpage}
\pagerange{\pageref{firstpage}--\pageref{lastpage}}
\maketitle

\begin{abstract}
We investigate the redshift evolution of disk and bulge-dominated galaxies using a mass-complete sample of $\sim$14,000 galaxies from the CANDELS survey, selected with $H_{\rm mag} \leq 24$, $M_{\rm stellar} \geq 10^9,{\rm M}_\odot$, and spanning $0.2 \leq z \leq 2.4$. Adopting an unbiased morphological classification, free from visual inspection or parametric assumptions, we explore the evolution of specific star formation rate (sSFR), stellar mass, structural properties, and galaxy fractions as a function of redshift and morphology. We find that while disk and bulge-dominated galaxies exhibit similar sSFR distributions at $z\sim2.4$, bulge-dominated systems develop a redshift-dependent bimodality below $z<1.6$, unlike the unimodal behaviour of disks. This bimodality correlates with stellar mass: bulge-dominated galaxies with lower sSFR are significantly more massive and exhibit higher Sérsic indices than their star-forming counterparts, despite having similar effective radii. Based on a Gaussian mixture decomposition, we identify two evolutionary tracks for bulge-dominated galaxies: G1, a long-lived, star-forming population with disk-like properties; and G2, a quenched, massive population whose prominence increases with decreasing redshift. The evolution of the star formation main sequence and morphology–mass fractions support a scenario in which G2 systems form through merger-driven transformations of massive disks. Our results indicate that bulge-dominated galaxies are not a homogeneous population, but instead follow divergent evolutionary paths driven by distinct physical mechanisms.
\end{abstract}




\section{Introduction}

Understanding the evolution of galaxies across cosmic time remains a challenge in astrophysics. Morphology encapsulates key information about the processes that shape galaxy formation and transformation \citep{2019igfe.book.....C}. A galaxy's morphology may originate from its initial angular momentum during halo collapse, with high angular momentum favoring disk-like structures and low angular momentum favoring more bulge-dominated galaxies \citep{1969ApJ...155..393P, 2015ApJ...812...29T}. However, after formation, galaxies undergo various physical processes -- such as major mergers \citep{2016MNRAS.458.2371R, 2018MNRAS.478.3994C, 2023MNRAS.518.5323H} and AGN feedback \citep{2013MNRAS.433.3297D, 2022MNRAS.509.3889D, 2023MNRAS.524.5327S} -- that can alter their star formation and structural properties, potentially triggering morphological transformations \citep[e.g.][]{2009ApJ...707..250M, 2013MNRAS.433.3297D}.

Galaxy morphology correlates strongly with star formation activity. In the local Universe, galaxies form a bimodal distribution in the star formation rate (SFR) versus stellar mass ($M_{\rm stellar}$) plane, known as the star formation main sequence (SFMS). This distribution distinguishes the blue cloud (actively star-forming galaxies, typically disk-dominated), the red sequence (quiescent galaxies, often bulge-dominated), and the transitional green valley \citep{2009MNRAS.396..818S, 2014MNRAS.440..889S, 2019MNRAS.488L..99A, 2022MNRAS.509..567S}. The co-evolution of morphology and star formation activity over cosmic time suggests that many galaxies transition from star-forming disk-dominated to quiescent bulge-dominated systems, with the specific evolutionary pathway shaped by the underlying physical mechanisms \citep[e.g.][]{2022MNRAS.509.3889D}.

This transition is also mass-dependent. Observations show that massive galaxies form their stars earlier and quench faster than their low-mass counterparts. \citep[``downsizing'',][]{2005ApJ...625..621B, 2006MNRAS.372..933N}. The peak of cosmic star formation density occurred at redshifts $z \sim 2-3$, when galaxies, regardless of morphology, were more active in forming stars \citep{2014ARA&A..52..415M, 2023MNRAS.519.1526P}. Concurrently, morphological fractions evolved, with a rising number of early-type galaxies at lower redshifts \citep{2007ApJS..172..494S, 2008ApJ...672..177L}.

Quantitative morphological classification is essential for tracing the evolution of galaxy populations over cosmic time. While visual classification remains intuitive, it becomes increasingly unreliable at high redshift due to surface brightness dimming and resolution limitations, which lead to significant discrepancies in the estimated fractions of different morphological types beyond $z = 1$ \citep{conselice2011tumultuous, mortlock2013redshift, talia2014listening, kartaltepe2015candels}. Parametric methods based on Sérsic profile fitting can also be misleading, particularly for irregular or clumpy galaxies, often yielding similar Sérsic indices for both disk and bulge-dominated systems \citep[e.g.][]{2022ApJ...938L...2F}.

As an alternative, non-parametric indices such as Concentration, Asymmetry, and Smoothness \citep[CAS;][]{conselice2003} provide morphological characterization without relying on assumed light profiles, though they remain sensitive to image preprocessing steps. Despite CAS being widely used in the literature, the combination of M20 \citep{lotz2004new} with Shannon entropy \citep{ferrari2015morfometryka, 2020A&C....3000334B} and gradient pattern asymmetry \citep{rosa2018gradient} -- MEGG system -- has demonstrated a better performance in separating elliptical and spiral galaxies in the local universe, as shown in \cite{Kolesnikov2024a} (see their Figs. 3 and 4).

In recent years, machine and deep learning approaches have enabled scalable, automated morphological classification, though these methods often rely on training sets derived from visual inspection and may inherit their biases \citep{dominguez2018improving, cheng2023lessons, walmsley2023galaxy}. \citet{2025MNRAS.tmp..593K} overcome this limitation by implementing a hybrid unsupervised-supervised method to classify galaxies from the CANDELS survey using redshift-binned models based on non-parametric MEGG indices, without any visual labeling.

In this work, we adopt the unbiased morphological classifications of \citet{2025MNRAS.tmp..593K} to investigate how disk- and bulge-dominated galaxies evolve across $0.2 \leq z \leq 2.4$. We analyze how specific star formation rate (sSFR), stellar mass, and SFR evolve for each morphological class. We identify two distinct bulge-dominated populations and explore their evolutionary pathways and structural properties. Finally, we use the evolution of morphological fractions as a function of redshift and stellar mass to constrain the processes leading to the formation of present-day massive ellipticals.

This paper is organized as follows: in Section 2, we present the sample and the catalogs we retrieve galaxy properties; in Section 3 we compare the evolution of disk and bulge-dominated galaxies as a function of redshift and estimate the merger rate; in Section 4 we focus on characterizing and exploring the evolution of two different families of bulge-dominated galaxies; in Section 5 we discuss our findings and compare them to previous results found in the literature; in Section 6, we summarize our main results. Throughout this paper we utilize a flat $\Lambda{\rm CDM}$ cosmology, with $[\Omega_{M}, \Omega_{\Lambda}, H_{0}] = [0.27, 0.73, 72 \, {\rm km \, s^{-1} \, Mpc^{-1}}]$ to be consistent with the results from the Planck Collaboration \citep{2016A&A...594A..13P} and report the magnitudes in the AB system.

\section{Data and sample selection}

We select galaxies from the Cosmic Assembly Near-infrared Deep Extragalactic Legacy Survey \citep[CANDELS;][]{faber2011cosmic, koekemoer2011candels, grogin2011candels}\footnote{Data catalogs were retrieved from the Mikulski Archive for Space Telescopes (MAST), available at \url{https://archive.stsci.edu/hlsp/candels}}. 
CANDELS is designed to investigate galaxy formation and evolution through deep near-infrared imaging of faint and distant galaxies, using the Wide Field Camera 3 (WFC3) aboard the \textit{Hubble Space Telescope} (HST). 
The survey includes observations of over 250{,}000 galaxies across five main fields, covering a total area of approximately $0.22 \, \mathrm{deg}^2$ \citep[e.g.][]{barro2019candels}: 
\begin{enumerate}
    \item the Cosmic Evolution Survey \citep[COSMOS;][]{2007ApJS..172....1S}, 
    \item the UKIDSS Ultra Deep Survey \citep[UDS;][]{2007MNRAS.379.1599L}, 
    \item the Extended Groth Strip \citep[EGS;][]{2007ApJ...660L...1D}, and 
    \item the Great Observatories Origins Deep Survey South and North fields \citep[GOODS-S and GOODS-N;][]{2004ApJ...600L..93G}. 
\end{enumerate}
Galaxy detection and photometric characterization in these fields are based on observations in the H-band (F160W; $\lambda_{\rm eff}^{\rm F160W} = 14{,}445$\,\AA).

Whereas COSMOS, UDS, EGS, and GOODS-S share similar photometric characterization and stellar mass estimation methods (see next subsection), the GOODS-N field follows a distinct procedure. For consistency, we exclude GOODS-N from our analysis. The initial sample contains 149{,}988 galaxies. Following previous studies indicating unreliable morphological classifications for sources fainter than $H_{\rm mag} = 24$ in the CANDELS fields \citep{grogin2011candels,kartaltepe2015candels}, we apply a magnitude cut, retaining 30{,}193 galaxies. We further restrict the sample to galaxies observed in both F160W (WFC3) and F814W (ACS), as the latter offers superior resolution (0.05\arcsec/pixel vs. 0.1\arcsec/pixel), which is crucial for morphological classification. This yields a final sample of 26{,}509 galaxies.

\subsection{Galaxy properties from Spectral Energy Distribution}

We retrieve $M_{\rm stellar}$ and SFR from \citet{santini2015stellar} and \citet{barro2019candels}, both based on Spectral Energy Distribution (SED) fitting. A major strength of the CANDELS fields is their extensive multi-wavelength coverage, with over 10 broadband filters spanning UV to mid-IR wavelengths. The bands adopted for each field are:
\begin{itemize}
    \item GOODS-S: 18 bands including UV (CTIO/MOSAIC U, VLT/VIMOS U), optical (HST/ACS F435W, F606W, F775W, F814W, F850LP), near-IR (HST/WFC3 F098M–F160W, VLT/ISAAC Ks, HAWK-I Ks), and mid-IR (Spitzer/IRAC 3.6–8.0\,$\mu$m);
    
    \item UDS: 19 bands including optical (CFHT/Megacam U; Subaru B, V, $R_c$, $i'$, $z'$), near-IR (VLT/HAWK-I Y, Ks; UKIDSS J, H, K), and the same HST and Spitzer coverage as GOODS-S;
    
    \item COSMOS: Similar to GOODS-S and UDS, with UV (GALEX), optical (Subaru B, V, $R_c$, $i'$, $z^+$), near-IR (HST/ACS F814W, WFC3 F125W, F160W; CFHT Ks), and mid-IR (Spitzer/IRAC 3.6–8.0\,$\mu$m);
    
    \item EGS: HST (ACS and WFC3), optical (Keck/DEIMOS B–Z; CFHT Legacy Survey), near-IR (UKIDSS J, H, K; WFC3 F125W, F160W), and mid-IR (Spitzer/IRAC 3.6–8.0\,$\mu$m).
\end{itemize}

The extensive photometric coverage enables robust stellar mass estimates. To assess systematic uncertainties, \citet{barro2019candels} conducted a cross-comparison where ten teams applied different SED fitting codes and assumptions—such as SFHs, extinction laws, and stellar population models—to the same photometric and redshift data. Most teams used \citet[BC03]{2003MNRAS.344.1000B} templates, with some including nebular emission, which is critical for young, star-forming galaxies where line emission affects near-IR fluxes. Stellar mass uncertainties are $\sim$30\% for $M_{\rm stellar} \geq 10^9\,{\rm M}_\odot$, and increase beyond 35\% at lower masses due to reduced signal-to-noise and greater model degeneracy. To ensure reliable estimates, we limit our sample to galaxies with $M_{\rm stellar} \geq 10^9\,{\rm M}_\odot$, resulting in 14{,}776 objects.

Hereafter, we adopt the $\mathrm{SFR_{UV,corr}}$ from \citet{barro2019candels}, which combines UV continuum measurements with dust extinction corrections based on the UV spectral slope (IRX–$\beta$ relation). This method provides good agreement with infrared-based estimates, particularly for galaxies with moderate dust content, yielding typical uncertainties of $\sim$0.3 dex. Although FIR-based SFRs (derived from Spitzer/MIPS 24\,$\mu$m and Herschel PACS/SPIRE data covering 24–500\,$\mu$m) offer more direct obscured SFR estimates—with uncertainties as low as 0.2 dex—they are available for only $\sim$5\% of our sample. To preserve statistical power across redshift bins, we therefore use $\mathrm{SFR_{UV,corr}}$ for the full sample. A detailed comparison between UV+IR and $\mathrm{SFR_{UV,corr}}$ estimates is presented in Appendix~\ref{sec:appendix_mstarsfr}.

\subsection{Characterizing morphology}

Although many studies have investigated galaxy evolution across a wide redshift range, the role of morphology remains incompletely understood. Defining galaxy morphology at high redshift is particularly challenging due to limited resolution, decreasing angular scale, and surface brightness dimming—all of which become increasingly significant with redshift. To avoid subjective visual inspection, we adopt a hybrid unsupervised-supervised algorithm based on non-parametric indices to classify galaxies between disk and bulge-dominated systems. We further detail bulge-dominated galaxies structural parameters using a Sérsic profile, namely the effective radius and Sérsic index.

\subsubsection{Assessing morphology through non-parametric indices}

In our first approach, we use non-parametric indices to characterize galaxy morphology.
Despite CAS being widely used in the literature, the combination of M20 \citep{lotz2004new} with Shannon entropy \citep{ferrari2015morfometryka, 2020A&C....3000334B} and gradient pattern asymmetry \citep{rosa2018gradient} -- MEGG system -- has demonstrated a better performance in separating elliptical and spiral galaxies in the local universe, as shown in \cite{Kolesnikov2024a} (see their Figs. 2 and 3). We therefore exclude CAS from our analysis and instead adopt the recently proposed MEGG system, which combines $M_{20}$, Shannon Entropy, Gini Index, and Gradient Field Asymmetry ($G_2$). This set of indices has demonstrated improved performance in distinguishing morphological types using both SDSS \citep{2000AJ....120.1579Y} and HST data \citep{1998SPIE.3356..234F}. Below we define each index:

\begin{itemize}
    \item \textbf{Second moment of light ($M_{20}$)}: Measures the spatial distribution of a galaxy's brightest regions. The total second-order moment is
    \begin{equation}
        M_{\rm tot} = \sum_i f_i \left[ (x_i - x_c)^2 + (y_i - y_c)^2 \right],
    \end{equation}
    where $f_i$ is the flux at pixel $(x_i, y_i)$, and $(x_c, y_c)$ is the galaxy center. $M_{20}$ is computed from the brightest 20\% of the total flux:
    \begin{equation}
        M_{20} = \log \left( \frac{\sum_i M_i}{M_{\rm tot}} \right), \quad \text{with } \sum f_i \leq 0.2 F_{\rm tot}.
    \end{equation}
    Lower $M_{20}$ values indicate concentrated light distributions (e.g., ellipticals); higher values reflect more extended structures (e.g., disks).

    \item \textbf{Shannon Entropy ($E$)}: Quantifies the flux distribution disorder. Given pixel fluxes $f_i$ over $N_{\rm p}$ pixels, with $p_i = f_i / \sum_j f_j$, the entropy is
    \begin{equation}
        E = -\sum_{i=1}^{K} p_i \log p_i,
    \end{equation}
    where $K$ is the number of bins. Bulge-dominated tend to have lower entropy, while disks show more uniform distributions.

    \item \textbf{Gini Index ($G$)}: Measures flux concentration across image pixels, ranging from 0 (uniform) to 1 (all flux in one pixel). Given ordered fluxes $f_i$,
    \begin{equation}
        G = \frac{1}{\bar{I} N_{\rm p}(N_{\rm p}-1)} \sum_{i=1}^{N_{\rm p}} (2i - N - 1) f_i,
    \end{equation}
    where $\bar{I}$ is the mean flux. High Gini values are typical of bulge-dominated; low values indicate disk-like morphologies.

    \item \textbf{Gradient Field Asymmetry ($G_2$)}: Based on Gradient Pattern Analysis, $G_2$ measures bilateral asymmetries in image gradients. The gradient field is computed across the image, assigning a vector (magnitude and direction) to each pixel. Vectors equidistant from the galaxy center are paired, and symmetric ones are discarded. The ``confluence'' $C$ quantifies vector alignment:
    \begin{equation}
        C = \left( \frac{\left|\sum_i v_a^i\right|}{\sum_i |v_a^i|} \right),
    \end{equation}
    where $v_a^i$ are asymmetric vectors. The $G_2$ index is then defined as
    \begin{equation}
        G_2 = \frac{V_{\rm A}}{V}(1 - C),
    \end{equation}
    where $V_{\rm A}$ is the number of asymmetric vectors and $V$ is the total number of pixels. This index effectively distinguishes early- and late-type galaxies, with demonstrated accuracy exceeding 90\% using SDSS data \citep{rosa2018gradient}.
\end{itemize}

\subsubsection{Defining disk and bulge-dominated systems through an hybrid method}
\label{subsec:hybrid}
Although non-parametric indices quantitatively capture galaxy morphology, their ability to separate classes (e.g., disks vs. bulge-dominated) becomes less reliable over broad redshift ranges. To address this, we adopt the classification from \citet{2025MNRAS.tmp..593K}, which employs a hybrid, eye-free method. In the first stage, a Self-Organizing Map \citep[SOM;][]{kohonen1991self} uses the non-parametric indices to group galaxies into clusters based on statistical similarity. Prominent clusters—those with distinct metric distributions—serve as labeled sets for a supervised deep learning stage.

The second stage uses a Convolutional Neural Network (CNN) based on the Xception architecture, trained to classify galaxies as disks or bulge-dominated. To improve reliability, 100 CNNs are trained on varying training-validation splits, and the final classification is taken as the average across the ensemble. Redshift evolution and morphological degradation are mitigated by using dedicated models for redshift bins spanning $0.2 \leq z \leq 2.4$ in steps of 0.2, as in \citet{2025MNRAS.tmp..593K}. The effect of redshift degradation is further quantified with the FERENGI code \citep{barden2008ferengi}, which simulates how low-redshift galaxies would appear at higher redshifts. These tests show that without correction, up to 18\% of disks could be misclassified as bulge-dominate -- highlighting the importance of this hybrid approach for recovering morphology across cosmic time.

Our final sample consists of $\sim$14{,}000 galaxies classified as ``disk-dominated'', ``bulge-dominated'', ``irregular'', with only a very small percentage ($<1\%$) of galaxies remain unclassified. To ensure robustness, we restrict our analysis to disk and bulge-dominated systems, excluding ambiguous cases. This results in a sample of 11{,}899 galaxies (7{,}479 disks and 4{,}420 bulge-dominated). Irregulars are only included when computing the total galaxy population per redshift bin, used in estimating the disk and bulge-dominated fractions.

\begin{figure*}
    \centering
    \includegraphics[width=\textwidth]{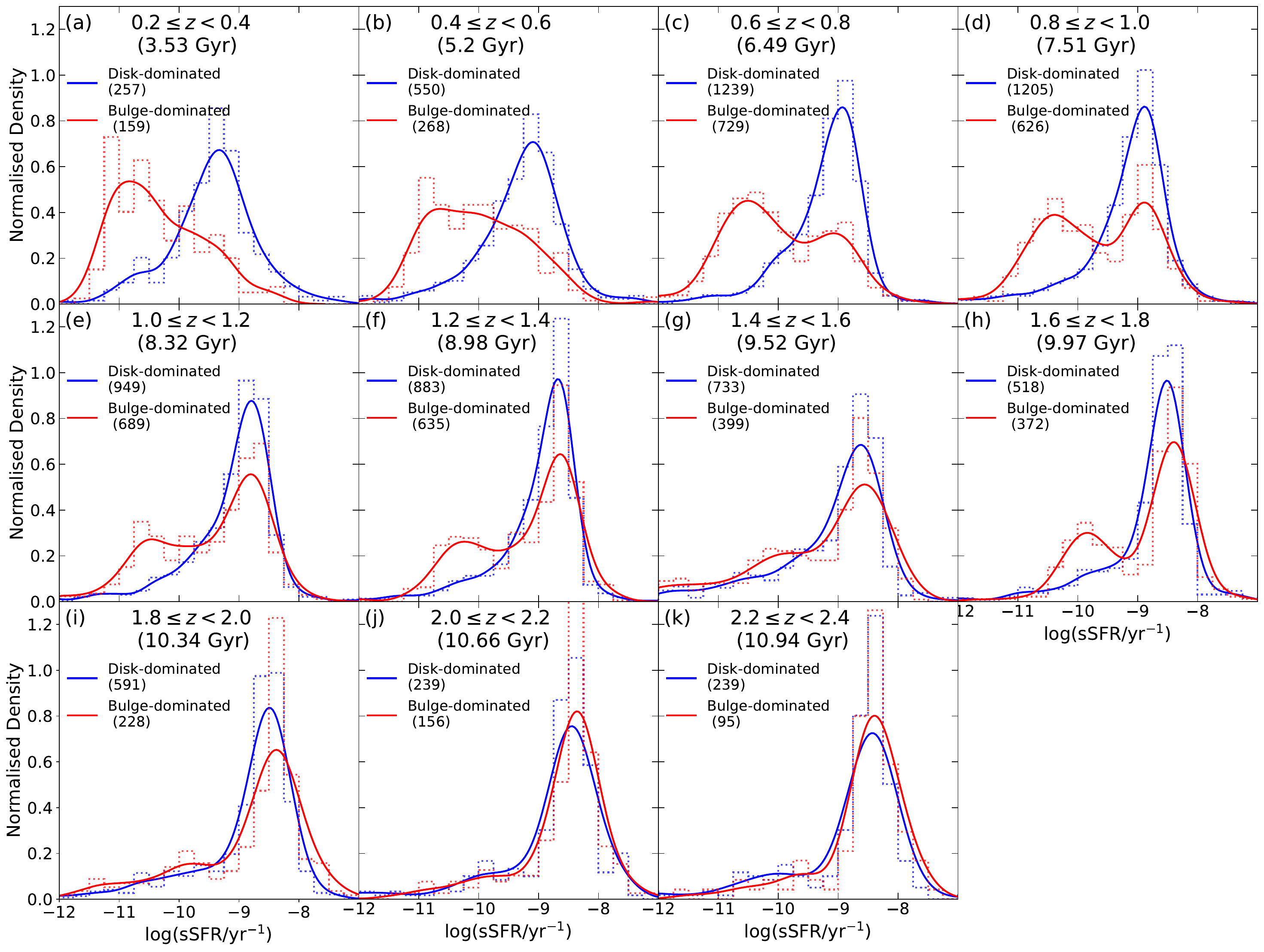}
    \caption{Distribution of sSFR for disk and bulge-dominated galaxies. Each panel shows the distribution for a given redshift bin. Dotted lines represent the histograms, whereas the solid lines show an adaptive kernel density estimate, applied to highlight overall trends. In each panel, we also include the number of galaxies in each morphological class in the legend. Despite starting from roughly the same sSFR distribution in panel (k), we note increasing differences towards lower redshifts. Bulge-dominated galaxies show a steeper decrement in their sSFR in comparison to disk galaxies, especially below $z < 1$.}
    \label{fig:s_d_ssfr}
\end{figure*}

\subsubsection{Describing light profiles with Sérsic fitting}

Elliptical galaxies are typically well described by a Sérsic profile, characterized by an effective radius ($R_{\rm eff}$)—the projected radius enclosing 50\% of the total light—and a Sérsic index ($n_{\rm s}$). We retrieve these parameters from the catalog of \citet{2024MNRAS.532.3747N}, which provides structural measurements across multiple CANDELS filters using \textsc{galfitm} \citep{2013MNRAS.430..330H, 2022A&A...664A..92H}, a multi-band extension of the widely used \textsc{galfit} software\footnote{\textsc{galfit} models galaxy light profiles using parametric functions, such as exponential laws for disks and Sérsic profiles for bulges \citep[e.g.][]{2013MNRAS.435..623V, 2014MNRAS.444.3603V, 2022A&A...664A..92H}.}. Following \citet{2024MNRAS.532.3747N}, we exclude galaxies with F160W signal-to-noise ratio (S/N) below 14, corresponding to a magnitude cut $\sim$2 mag brighter than the 90\% completeness limit. Although this cut is applied in the F160W band, we adopt their structural measurements obtained in the F814W filter. We highlight that the detailing of galaxies' morphology through Sérsic index and effective radius is limited to galaxies labeled as bulge-dominated according to the method described in \ref{subsec:hybrid}.

Uncertainties in $R_{\rm eff}$ and $n_{\rm s}$ depend on factors such as S/N, galaxy brightness, and structural complexity. For bright galaxies, typical $R_{\rm eff}$ uncertainties are $\sim$5–10\%, but can exceed 20\% for fainter systems, especially in low S/N data gathered with the F814W filter. To ensure reliability, we discard sources where the uncertainty in $R_{\rm eff}$ exceeds 20\%.

\section{Comparing disks and bulge-dominated galaxies}

Disks and bulge-dominated systems follow different pathways regarding formation and evolution. The former is expected to be the result of gas cloud collapsing with high angular momentum, whereas the second is formed in the case of negligible angular momentum. After formation, several mechanisms can alter a galaxy morphology. Therefore, it is fundamental to investigate the evolution of both classes separately.

\subsection{Evolution in specific star formation}

We begin by analyzing the evolution of the specific star formation rate (sSFR) for disk and bulge-dominated galaxies across redshift, as shown in Fig.~\ref{fig:s_d_ssfr}. The sSFR provides a measure of galaxies growth in stellar mass relative to its current stellar mass, thus being a tracer of the system evolutionary stage. Each panel presents the sSFR distribution for a given redshift bin, with disk and bulge-dominated galaxies shown in blue and red, respectively. Dotted lines denote histograms of the observed data, while solid lines show adaptive kernel density estimates to highlight trends. The number of galaxies in each morphological class and the corresponding look-back time are indicated.

At high redshift ($z \sim 2.3$), the sSFR distributions of disks and bulge-dominated galaxies largely overlap, indicating that both populations were actively forming stars. However, toward lower redshifts, the two populations diverge markedly. Disk galaxies display a gradual decline in sSFR, with a median change of $\Delta {\rm sSFR}_{\rm 2.3-0.3}^{\rm Disks} \sim -0.86$ dex. Their sSFR distribution remains platycurtic (i.e., heavy-tailed relative to a Gaussian), suggesting a persistent population of lower-sSFR disks throughout cosmic time.

Bulge-dominated galaxies, in contrast, exhibit a sharper drop in sSFR, with $\Delta {\rm sSFR}_{\rm 2.3-0.3}^{\rm Bulge-Dominated} \sim -2.05$ dex. While their distribution initially resembles that of disks, a pronounced bimodality emerges at $z \lesssim 1.3$, becoming dominant below $z \sim 0.7$. From this point onward, the sSFR distribution splits into two components: (1) a high-sSFR component similar to disks, and (2) a quenched component with significantly lower sSFR values. This growing dominance of the low-sSFR mode suggests that star formation in bulge-dominated galaxies is increasingly suppressed with cosmic time. The transition is also visible as a significant shift in the distribution mode, hinting at different evolutionary channels leading to the formation of bulge-dominated systems.

The observed divergence in sSFR evolution between disks and bulge-dominated galaxies supports the idea that star formation quenching proceeds more rapidly in bulge-dominated systems, consistent with prior studies showing accelerated evolutionary timescales for compact or bulge-dominated galaxies \citep[e.g.,][]{2013ApJ...765..104B, 2015ApJ...811L..12W}. The presence of a bimodal sSFR distribution in bulge-dominated systems at $z \lesssim 1.3$, and its growing dominance toward lower redshifts, suggests that not all bulge-dominated systems are passively evolving remnants of early-type galaxies. Instead, a significant fraction of them may still be in transition, or may have formed via distinct evolutionary pathways. This duality is in agreement with results from \citet{2017MNRAS.472.2054P}, who find that the quiescent population can arise from both fast-track and slow-track quenching mechanisms.

The emergence of this bimodality raises the possibility that the morphological transformation into a bulge-dominated system may be linked—though not universally—to star formation quenching. Since sSFR is sensitive to both stellar mass and SFR, we next investigate whether this bimodality is driven primarily by a change in SFR, stellar mass, or both.

\subsection{Star formation rate, stellar mass, or both?}
\label{subsec:sfr_mstellar_both}

\begin{figure*}
    \centering
    \includegraphics[width=\textwidth]{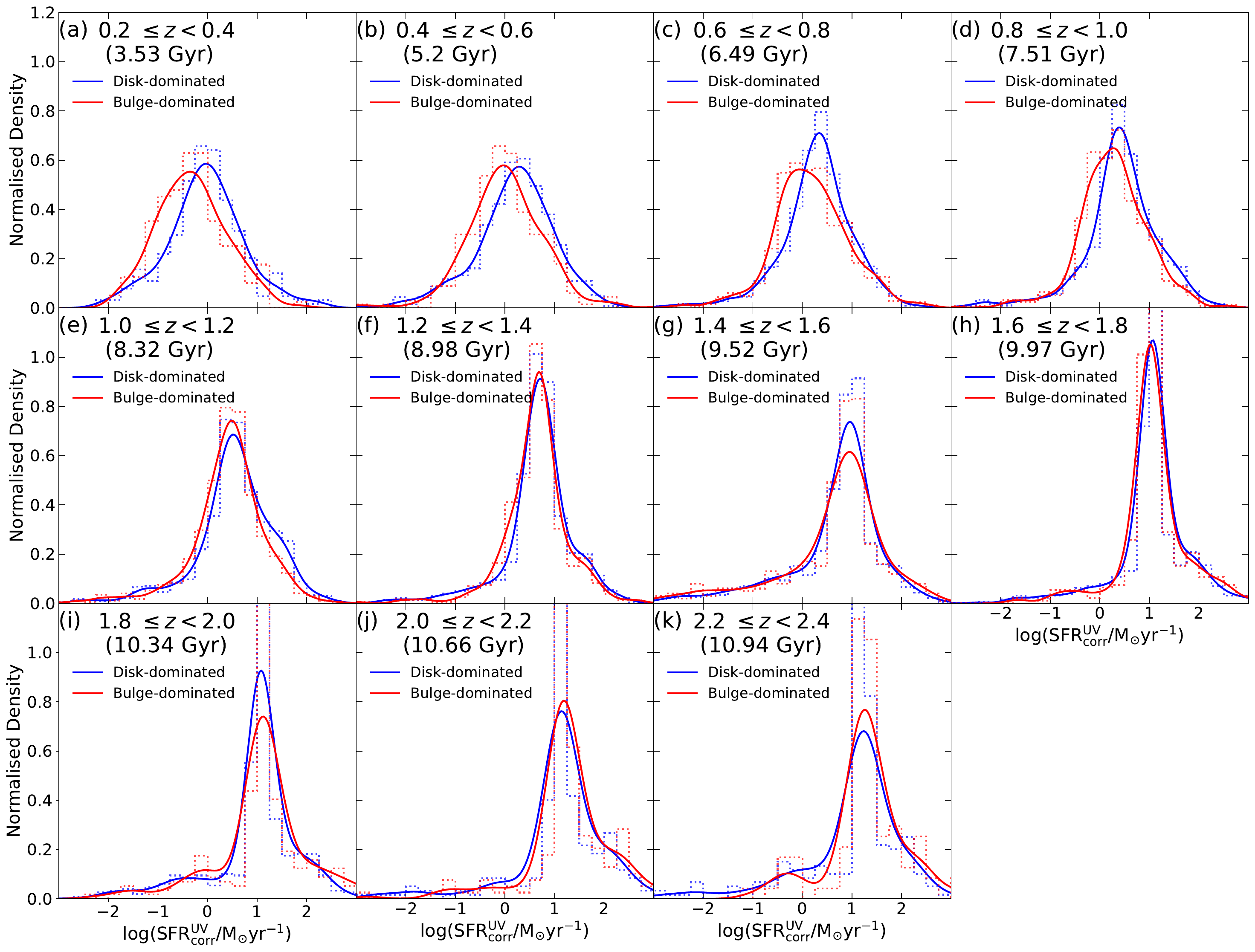}
    \caption{Analogue to Fig.~\ref{fig:s_d_ssfr}, but for the SFR.}
    \label{fig:s_d_sfr}
\end{figure*}

To disentangle the physical drivers of the bimodality observed in the sSFR distribution of bulge-dominated galaxies, we analyze the distributions of SFR and stellar mass separately. Figure~\ref{fig:s_d_sfr} shows that the SFR distributions of disks and bulge-dominated systems are remarkably similar across all redshifts, peaking at $\log({\rm SFR}/{\rm M}_{\odot}\,{\rm yr}^{-1}) \sim 1.2$ at $z \sim 2.3$, with extended tails toward lower SFR. These tails explain the platycurtic nature of the sSFR distribution seen in disks. Even at high redshift, a non-negligible fraction of both disks (14\%) and bulge-dominated galaxies (11\%) have low SFRs ($\log{\rm SFR} \leq 0$), suggesting early signs of star formation suppression. With cosmic time, the SFR distributions become more symmetric and gradually shift to lower values. Notably, the average SFR difference between disks and bulge-dominated galaxies remains small—$\sim$0.3 dex at $z \sim 0.2$—in contrast to the $>1$ dex separation seen in the local Universe \citep[e.g.][]{2022MNRAS.509..567S}. This suggests that the bulk of SFR quenching in bulge-dominated systems likely occurs at $z < 0.3$, within a $\sim$2 Gyr window, enabling a rapid transition from the green valley to the red sequence, consistent with quenching timescales of $\lesssim$2 Gyr, in agreement with, e.g., \citet{2013ApJ...765..104B, 2022ApJ...927..170T}.

\begin{figure*}
    \centering
    \includegraphics[width=\textwidth]{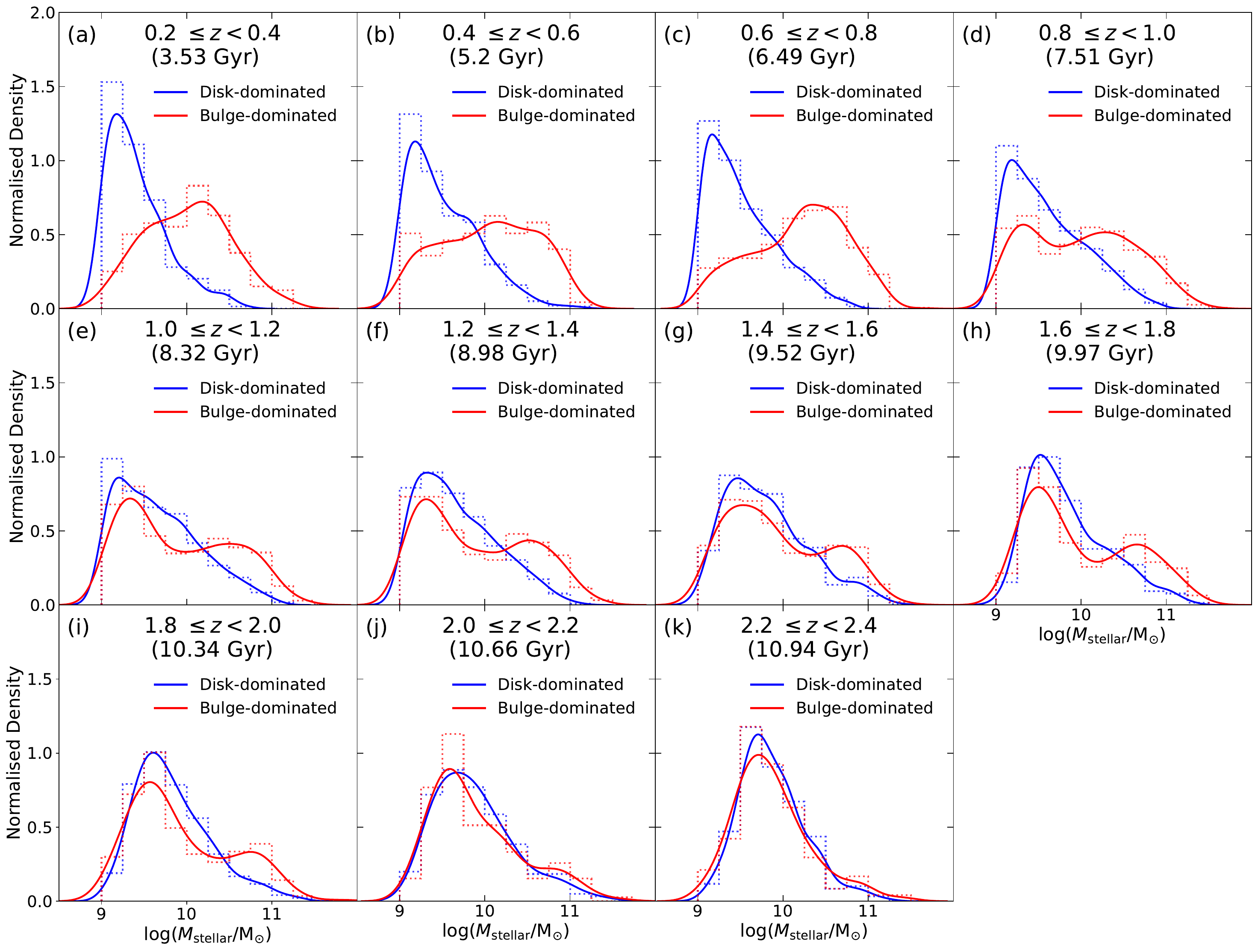}
    \caption{Similar to Fig.~\ref{fig:s_d_ssfr}, but for $M_{\rm stellar}$.}
    \label{fig:s_d_mstellar}
\end{figure*}

However, this marginal SFR difference alone cannot account for the pronounced divergence in sSFR. Figure~\ref{fig:s_d_mstellar} reveals that the origin of the bimodality lies in the evolution of stellar mass. While disks and bulge-dominated galaxies share nearly identical mass distributions at $z \sim 2.3$, their evolutionary paths diverge significantly. Disk galaxies maintain a moderately asymmetric mass distribution with a persistent high-mass tail across redshifts, consistent with a secular buildup of mass through inside-out star formation and gradual accretion \citep[e.g.,][]{2013ApJ...771L..35V, 2016ApJ...828...27N}. In contrast, bulge-dominated galaxies develop a distinct bimodality in their stellar mass distribution by $z \sim 1.7$, with a secondary, high-mass component becoming dominant below $z \sim 0.7$. The mode of the mass distribution for bulge-dominated galaxies increases by nearly 1 dex between $z \sim 0.9$ and $z \sim 0.7$, suggesting a rapid growth in the massive population.

This bimodality strongly suggests that the declining sSFR in bulge-dominated galaxies is driven primarily by stellar mass assembly rather than by SFR suppression alone. We suggest a scenario in which massive disk galaxies undergo morphological transformation—likely through major mergers—into bulge-dominated systems, a process that simultaneously increases stellar mass and steepens the light profile \citep[e.g.,][]{2006ApJS..163....1H, 2009ApJS..181..135H}. This scenario is supported by the increasing prominence of a high-mass bulge-dominated galaxies and the concurrent decline in massive disks, pointing toward a mass-dependent transition mechanism. The resulting downsizing trend, in which low-mass disks continue forming stars while high-mass systems quench and transform, naturally emerges from this picture \citep{2005ApJ...633..174T, 2018ApJ...854...30P}.

An alternative to explain the bimodality is the passive fading of asymmetric star-forming regions, that could cause disk galaxies to appear more bulge-dominated after quenching \citep[e.g.,][]{2011ApJ...743...87W, 2012ApJ...753..167B}. However, fading alone typically results in modest changes to a galaxy’s light profile and is unlikely to account for the observed steepening in Sérsic index ($> 4$) or the emergence of a distinct high-mass mode in the stellar mass distribution \citep{2012MNRAS.427.1666B}. In contrast, major mergers are known to efficiently destroy disk structures and redistribute stellar mass, producing concentrated bulge-dominated galaxies consistent with our observations \citep{2009ApJS..181..135H, 2009ApJ...699L.178N}. Moreover, the build-up of stellar mass in bulge-dominated systems may also involve the accretion of previously quenched satellites, further contributing to the growth of the high-mass population and reinforcing the bimodal mass distribution \citep{2010ApJ...725.2312O, 2016MNRAS.458.2371R}.

If this transformation is indeed mass-dependent, as suggested by the increasing dominance of massive bulge-dominated systems, one would expect the fraction of bulge-dominated galaxies to evolve strongly with redshift—particularly at the high-mass end. In the next section, we test this hypothesis by quantifying the evolution of disk and bulge-dominated fractions across different stellar mass bins. This provides an independent check on the proposed link between mass assembly, morphology transformation, and galaxy quenching.

\subsection{Signs of morphological transition}

\begin{figure*}
    \centering
    \includegraphics[width=\textwidth]{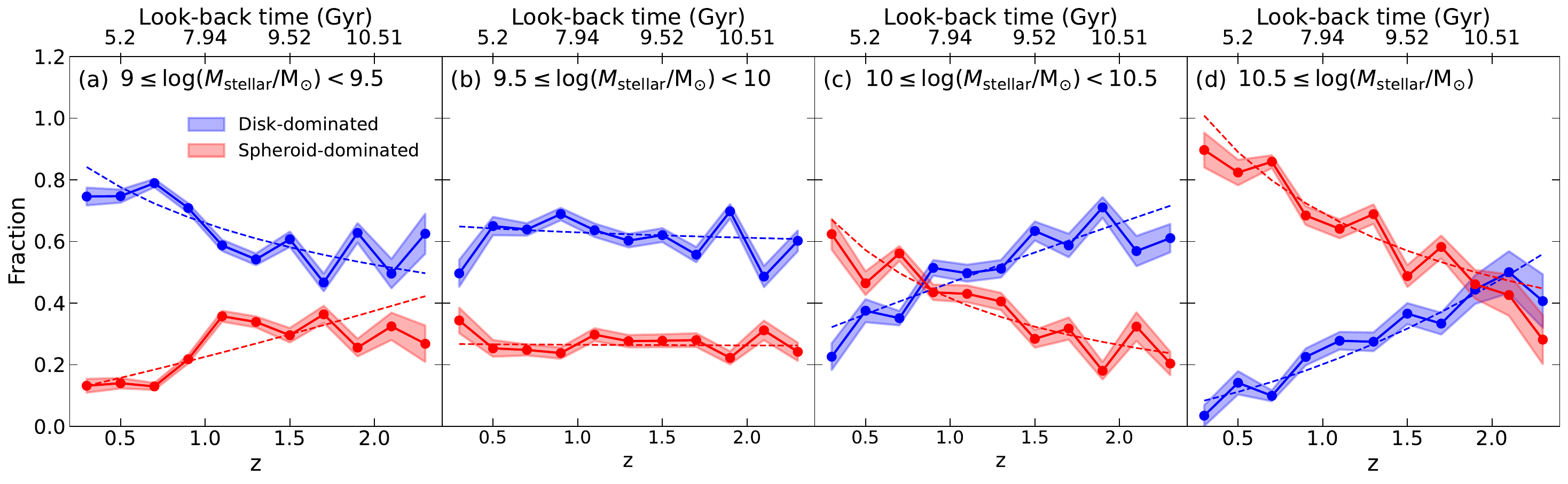}
    \caption{Fraction of disk (blue colored) and bulge-dominated (red colored) galaxies as a function of redshift, separating systems into four stellar mass bins. Shaded areas denote the uncertainty in the fraction of each bin, and is calculated by assuming a multinomial distribution (Equation~\ref{eq:multinomial_std}). We also show as dashed lines the results of fitting Equation~\ref{eq:fraction_form} to observed data.}
    \label{fig:s_d_fraction}
\end{figure*}

If massive disk galaxies evolve into bulge-dominated systems through mergers, as suggested by our results, we expect to observe corresponding changes in the morphological fractions of disks and bulge-dominated systems over cosmic time, particularly as a function of stellar mass. Figure~\ref{fig:s_d_fraction} shows the redshift evolution of the fraction of disk (blue) and bulge-dominated (red) galaxies across four stellar mass bins. Fractions are computed with respect to the total number of galaxies in each bin (including irregulars), and uncertainties are estimated using a multinomial distribution:
\begin{equation}
    \label{eq:multinomial_std}
    \sigma^{\rm Multinomial}_{\rm i} = \sqrt{n p_{\rm i} (1-p_{\rm i})},
\end{equation}
where $p_{\rm i}$ is the fraction of morphological type $i$, and $n$ is the total number of galaxies in the bin. To quantify trends, we fit each distribution using a power-law model:
\begin{equation}
    \label{eq:fraction_form}
    F_{\rm i}(z) = C_{\rm i} \times (1+z)^{m_{\rm i}}.
\end{equation}
The results are shown in Table \ref{tab:fraction_fitting}.

\begin{table}
\caption{Results of fitting equation \ref{eq:fraction_form} to the observed fractions of disk or bulge-dominated galaxies, separating into four different stellar mass bins.}
\label{tab:fraction_fitting}
\resizebox{\columnwidth}{!}{%
\begin{tabular}{ccc}
\hline
\multicolumn{3}{c}{Disk galaxies}                                                                                \\ \hline
\multicolumn{1}{c|}{Stellar Mass Bin}                                       & $C_{\rm d}$     & $m_{\rm d}$      \\ \hline
\multicolumn{1}{c|}{$9 \leq \log(M_{\rm stellar}/{\rm M}_{\odot}) < 9.5$}   & $0.98 \pm 0.06$ & $-0.57 \pm 0.09$ \\
\multicolumn{1}{c|}{$9.5 \leq \log(M_{\rm stellar}/{\rm M}_{\odot}) < 10$}  & $0.65 \pm 0.03$ & $-0.07 \pm 0.02$ \\
\multicolumn{1}{c|}{$10 \leq \log(M_{\rm stellar}/{\rm M}_{\odot}) < 10.5$} & $0.26 \pm 0.05$ & $0.86 \pm 0.05$  \\
\multicolumn{1}{c|}{$10.5 \leq \log(M_{\rm stellar}/{\rm M}_{\odot})$}      & $0.05 \pm 0.04$ & $2.04 \pm 0.03$  \\ \hline
\multicolumn{3}{c}{Bulge-dominated galaxies}                                                                          \\ \hline
\multicolumn{1}{c|}{Stellar Mass Bin}                                       & $C_{\rm s}$     & $m_{\rm s}$      \\ \hline
\multicolumn{1}{c|}{$9 \leq \log(M_{\rm stellar}/{\rm M}_{\odot}) < 9.5$}   & $0.09 \pm 0.03$ & $1.25 \pm 0.09$  \\
\multicolumn{1}{c|}{$9.5 \leq \log(M_{\rm stellar}/{\rm M}_{\odot}) < 10$}  & $0.27 \pm 0.04$ & $-0.02 \pm 0.03$ \\
\multicolumn{1}{c|}{$10 \leq \log(M_{\rm stellar}/{\rm M}_{\odot}) < 10.5$} & $0.90 \pm 0.07$ & $-0.72 \pm 0.08$ \\
\multicolumn{1}{c|}{$10.5 \leq \log(M_{\rm stellar}/{\rm M}_{\odot})$}      & $1.26 \pm 0.5$  & $-0.87 \pm 0.04$ \\ \hline
\end{tabular}%
}
\end{table}

We find strong mass-dependent trends. At low stellar masses ($\log M_\star/M_\odot < 10$), the fractions of disks and bulge-dominated systems remain relatively constant, with disks dominant in all redshift bins. The lowest-mass bin even shows a slight decrease in the bulge-dominated fraction below $z \sim 1$, possibly indicating late-stage disk regrowth or a more extended morphological transformation timescale in low-mass systems. In contrast, at higher masses ($\log M_\star > 10$), disk and bulge-dominated fractions evolve in opposite directions. For $\log M_\star/M_\odot > 10.5$, the fraction of disk galaxies falls from 55\% at $z \sim 2.4$ to 5\% at $z \sim 0.2$, while the bulge-dominated fraction increases from 25\% to 90\%. These symmetric and steep trends suggest that massive disks are progressively transformed into bulge-dominated galaxies, most likely through major mergers, as discussed in prior works \citep[e.g.,][]{2005ApJ...622L...9S, 2006ApJ...652..270B, 2009ApJS..181..135H, 2013MNRAS.428..999P}.

The nearly static morphological fractions observed in intermediate mass bins (e.g., $9.5 \leq \log M_\star < 10$) help explain why some studies report a weak or non-evolving morphological mix over time \citep[e.g.,][]{2025MNRAS.tmp..593K}. However, this stability is a result of stellar mass selection: galaxies in this mass range dominate the global number counts, especially at intermediate redshifts. These findings emphasize that morphological fraction trends are sensitive to the chosen mass range, and interpreting evolution without accounting for stellar mass may obscure underlying transitions—especially in studies using JWST or deep surveys \citep[e.g.,][]{ferrari2015morfometryka, dominguez2018improving, walmsley2022galaxy, cheng2023lessons}.

Altogether, these results provide evidence for a mass-dependent morphological transformation, where massive disk galaxies gradually disappear and are replaced by a growing population of massive bulge-dominated systems. The observed evolution is consistent with a merger-driven origin, and the symmetry between the decline in disks and rise in bulge-dominated galaxies suggests that this is not merely due to passive fading or observational bias. In the next section, we explore this scenario in greater detail by dissecting the bulge-dominated population into two subgroups—one with low stellar mass and high sSFR, and another with high mass and quenched star formation—to assess whether distinct evolutionary paths are encoded in their properties.

\section{Characterizing different evolutionary pathways of bulge-dominated galaxies}

The evolution observed in the specific star formation rate (sSFR) and stellar mass distributions (Figs.~\ref{fig:s_d_ssfr} and \ref{fig:s_d_mstellar}) suggests that bulge-dominated galaxies do not form or evolve as a single, homogeneous population. Instead, we identify two distinct subgroups that likely follow different evolutionary tracks. The first group—hereafter referred to as \textit{high-sSFR bulge-dominated galaxies}—comprises bulge-dominated galaxies with sSFR, SFR, and stellar mass distributions broadly consistent with those of disk galaxies. These systems are present across all redshifts and appear to be still forming stars, despite their bulge-dominated morphology. The second group—\textit{low-sSFR bulge-dominated galaxies}—includes galaxies with similar star formation rates to disk systems but significantly higher stellar masses, typically exceeding those of both disks and high-sSFR bulge-dominated galaxies by more than 1 dex. In this section, we isolate and characterize each subgroup in order to quantify their respective roles in the overall evolution of the bulge-dominated population.

\subsection{Characterizing bimodal distributions}

Multimodal distributions are a longstanding challenge in statistics, especially when evolving across a secondary variable like redshift. In our case, the sSFR distribution of bulge-dominated galaxies exhibits visual signs of bimodality, particularly at $z < 1.8$ (see Fig.~\ref{fig:s_d_ssfr}). To quantify this behavior, we decompose the sSFR distribution into multiple Gaussian components using a Gaussian Mixture Model (GMM), which provides a probabilistic framework to separate overlapping populations.

We perform the decomposition in each redshift bin using both frequentist and Bayesian approaches via \texttt{GaussianMixture} and \texttt{BayesianGaussianMixture} from \texttt{scikit-learn}, finding consistent results. To select the optimal number of components, we compute the Akaike Information Criterion (AIC) and Bayesian Information Criterion (BIC) for GMM fits ranging from 1 to 10 components (details in Appendix~\ref{sec_app:GMM}). Both criteria indicate that a two-component model best describes the data in the redshift range $z < 1.6$.

\begin{figure*}
    \centering
    \includegraphics[width=\textwidth]{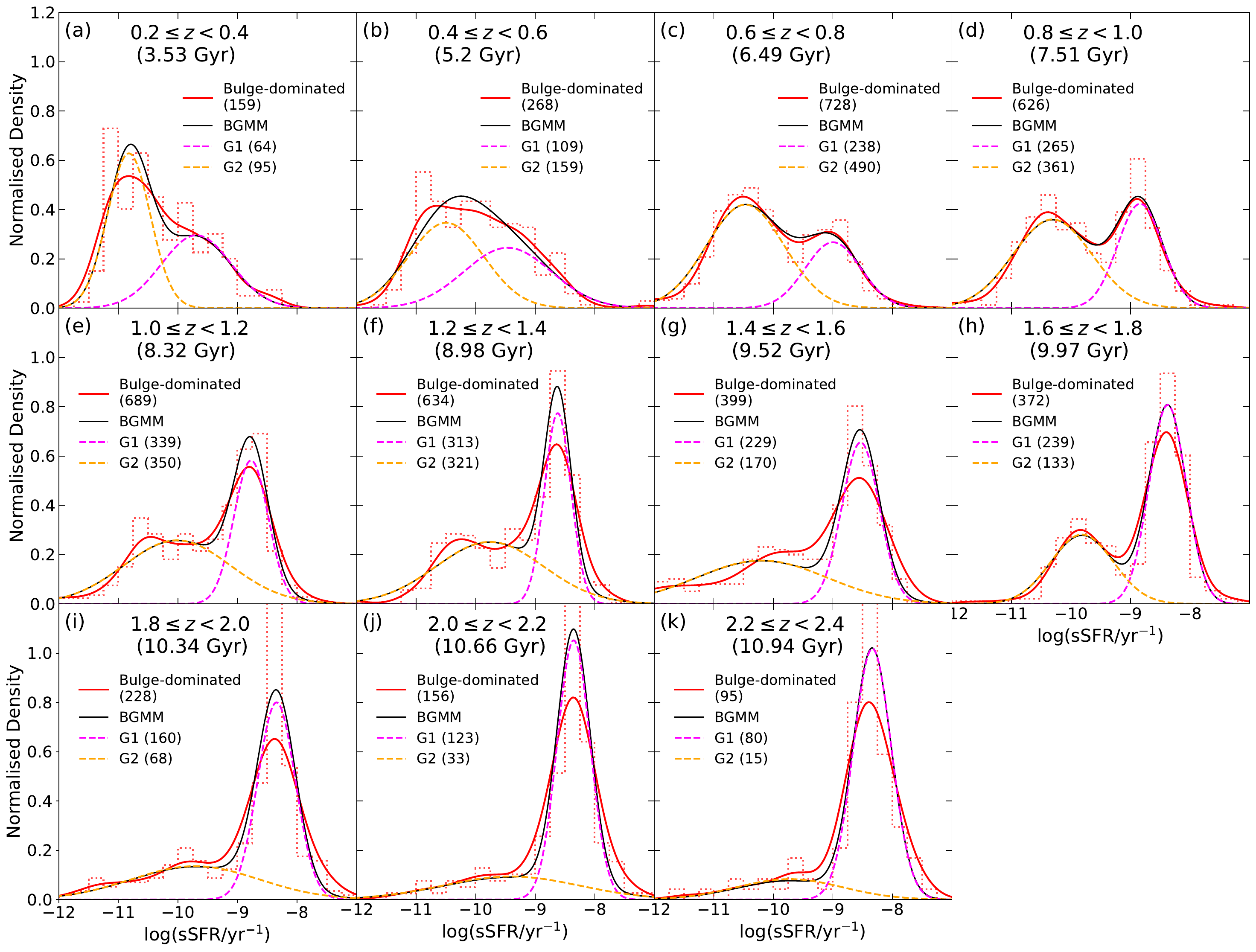}
    \caption{Decomposition of the observed sSFR distribution of bulge-dominated galaxies into two Gaussian components, with each panel representing a different redshift bin. Background histogram in red shows the observed distributions, the red solid line shows an adaptive kernel density applied to the data, the solid black line represents the resulting distribution from the BGMM method, whereas the magenta and orange dashed lines indicates the two different components found in the fit.}
    \label{fig:bimodality_ssfr}
\end{figure*}

Figure~\ref{fig:bimodality_ssfr} shows the resulting two-component decomposition for each redshift bin. The observed histogram of bulge-dominated galaxies is plotted in red, overlaid with an adaptive kernel density estimate (solid red line), the total GMM fit (solid black line), and the individual Gaussian components (dashed magenta and orange lines). We label the higher-sSFR component as G1 (magenta) and the lower-sSFR component as G2 (orange). The number of galaxies associated with each component is shown in the legend of each panel. As expected, the decomposition is most robust in the $z < 1.6$ regime, where bimodality is more visually apparent.

\begin{figure}
    \centering
    \includegraphics[width=0.8\columnwidth]{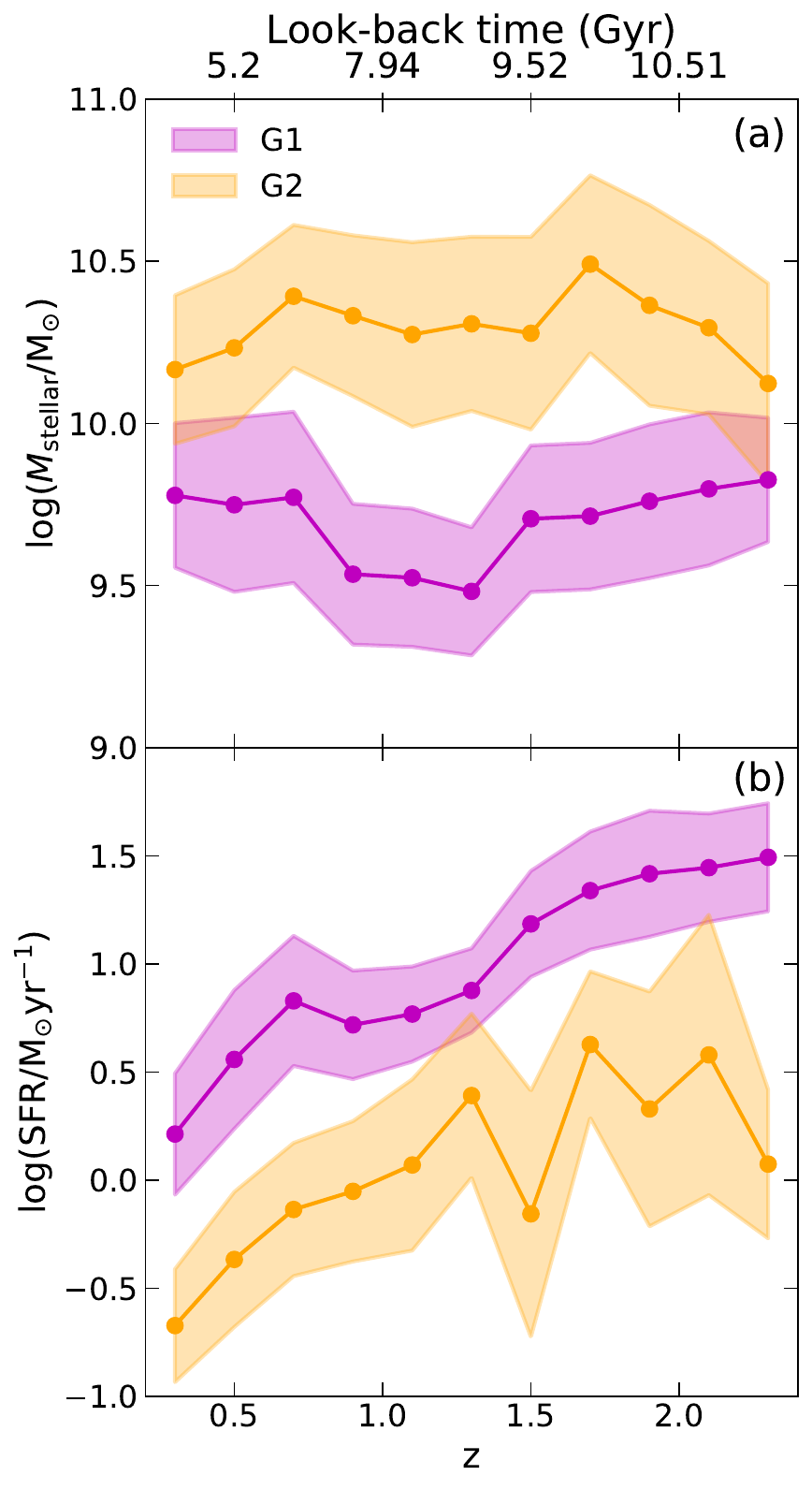}
    \caption{Evolution of mean stellar mass (panel a) and star formation rate (panel b) as a function of redshift for components G1 (magenta) and G2 (orange).}
    \label{fig:spheroid_bimodality_sfr_mstellar}
\end{figure}

To understand this transformation in terms of physical properties, we examine the stellar mass and SFR of each component (Fig.~\ref{fig:spheroid_bimodality_sfr_mstellar}). On average, G2 galaxies are more massive than G1 by $\Delta \log(M_{\rm stellar}/{\rm M}_\odot) = 0.61 \pm 0.16$ dex, while having lower SFRs by $\Delta \log({\rm SFR}/{\rm M}_\odot\,{\rm yr}^{-1}) = 0.92 \pm 0.26$ dex. Notably, the average stellar mass of G2 changes little with redshift ($\sim$0.3 dex), suggesting that this population is not growing primarily through in-situ star formation. Moreover, the standard deviation of the G2 sSFR component is significantly broader than that of G1 ($\langle \sigma_{2} \rangle = 0.81 \pm 0.24$ vs. $\langle \sigma_{1} \rangle = 0.38 \pm 0.14$), indicating greater heterogeneity in the quenching stage or recent assembly history. The consistent offset between the mean sSFR of the two components ($\langle \Delta \mu_{\rm G1-G2} \rangle = 1.31 \pm 0.19$ dex) further supports a scenario in which these represent distinct evolutionary phases.

\subsection{Star formation main sequence diagram}

\begin{figure*}
    \centering
    \includegraphics[width=\textwidth]{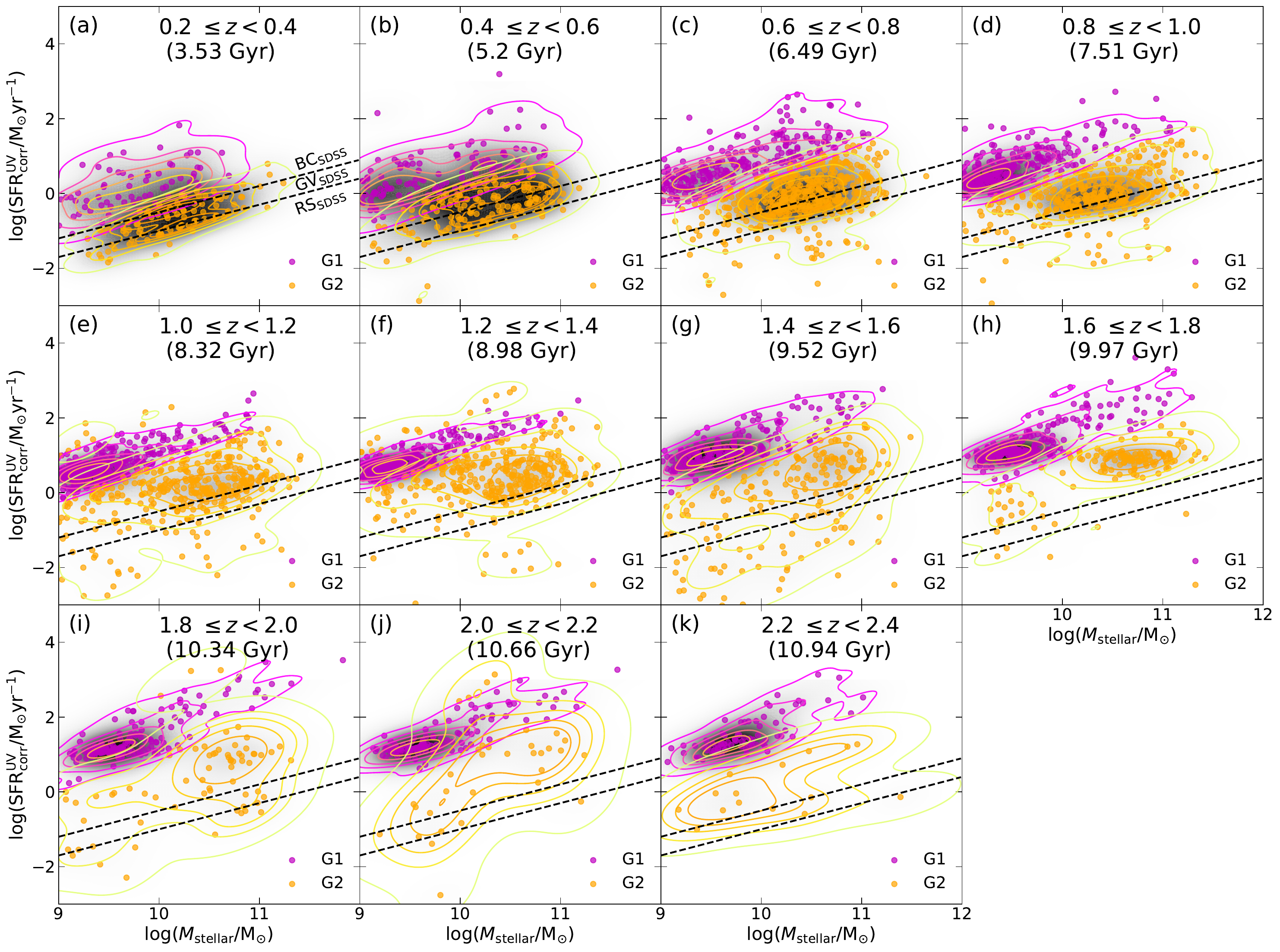}
    \caption{Distributions of G1 (magenta) and G2 (orange) in the SFMS diagram. Whereas scatter points represent the observed data, the contour lines (also in magenta and orange) traces the data density distribution as a function of location in the diagram. For completeness, and to create a reference for quantifying trends, we include as black dashed lines the local Universe based separation between BC, GV and RS, given by equations~\ref{eq:BC_GV}~and~\ref{eq:GV_RS}. Each panel represents a different redshift bin. The contours for G1 and G2 represent always the same levels.}
    \label{fig:bimodality_sfms}
\end{figure*}

To examine the evolutionary states of the two bulge-dominated subpopulations identified via GMM (G1 and G2), we analyze their locations in the star formation main sequence (SFMS) diagram. In Fig.~\ref{fig:bimodality_sfms}, we show the SFR–$M_{\rm stellar}$ distribution of G1 (magenta) and G2 (orange) for all redshift bins. Individual galaxies are shown as scatter points, and density contours emphasize structural differences in their distributions. For reference, we include dashed black lines marking the approximate boundaries between the blue cloud (BC), green valley (GV), and red sequence (RS), defined for the local Universe following \citet{2020MNRAS.491.5406T}:
\begin{align}
\log({\rm SFR}/{\rm M}_\odot\,{\rm yr}^{-1}) &= 0.7 \log(M_{\rm stellar}/{\rm M}_\odot) - 7.5 \label{eq:BC_GV} \\
\log({\rm SFR}/{\rm M}_\odot\,{\rm yr}^{-1}) &= 0.7 \log(M_{\rm stellar}/{\rm M}_\odot) - 8.0 \label{eq:GV_RS}
\end{align}
These divisions serve as a qualitative reference; they may shift with redshift, particularly the BC normalization \citep[e.g.,][]{2012ApJ...754L..29W, 2014ApJS..214...15S, 2016ApJ...817..118T}.

The G1 population consistently occupies the BC across all redshifts, indicating sustained star formation activity. In contrast, G2 galaxies show clear evolutionary progression: at $z \gtrsim 1.8$, they largely coincide with G1 in the BC, but toward lower redshifts, they shift into the GV and RS. This trend reflects the gradual quenching of the G2 population. n the lowest redshift bin ($z \sim 0.3$), only 17\% of G2 galaxies remain in the blue cloud, while 49\% are found in the green valley and 34\% have already transitioned to the red sequence.

To quantify this transition, we perform linear fits to the G1 and G2 distributions in the SFMS at each redshift (Table~\ref{tab:linear_fit_sfms}). For G1, the slope increases by $\sim 0.3$ from low to high redshift, while the intercept decreases—suggesting that massive, star-forming bulge-dominated galaxies are more common at earlier times. G2 exhibits non-linear and irregular behavior, as expected for a population in transition. For this reason, the fit parameters for G2 are included for completeness but should be interpreted cautiously.

\begin{table}
\caption{Linear fit results for the G1 and G2 distributions in the SFMS, for each redshift bin. We highlight that, despite we present the results for the G2 component for completeness, they should be analyzed carefully, as the G2 relation between star formation and stellar mass are highly non-linear. Column (1) show the redshift bin considered; columns (2) and (3) show the slope for G1 and G2 distributions, respectively; and columns (4) and (5) show the intercept for the same distributions.}
\label{tab:linear_fit_sfms}
\resizebox{\columnwidth}{!}{%
\begin{tabular}{c|cc|cc}
\hline
          & \multicolumn{2}{c|}{Slope}         & \multicolumn{2}{c}{Intercept}        \\ \hline
z bin     & G1               & G2              & G1               & G2                \\ \hline
0.2 - 0.4 & $0.73 \pm 0.13$  & $0.92 \pm 0.07$ & $-6.94 \pm 0.25$ & $-9.98 \pm 0.21$  \\
0.4 - 0.6 & $0.78 \pm 0.09$  & $0.81 \pm 0.08$ & $-7.03 \pm 0.16$ & $-8.70 \pm 0.15$  \\
0.6 - 0.8 & $0.89 \pm 0.05$  & $0.61 \pm 0.06$ & $-7.88 \pm 0.11$ & $-6.48 \pm 0.10$  \\
0.8 - 1.0 & $0.88 \pm 0.05$  & $0.67 \pm 0.06$ & $-7.63 \pm 0.12$ & $-6.95 \pm 0.10$  \\
1.0 - 1.2 & $0.86 \pm 0.03$  & $0.50 \pm 0.07$ & $-7.39 \pm 0.11$ & $-5.05 \pm 0.09$  \\
1.2 - 1.4 & $0.87 \pm  0.03$ & $0.29 \pm 0.08$ & $-7.37 \pm 0.12$ & $-2.57 \pm 0.10$  \\
1.4 - 1.6 & $0.88 \pm 0.04$  & $1.08 \pm 0.12$ & $-7.37 \pm 0.13$ & $-11.23 \pm 0.12$ \\
1.6 - 1.8 & $1.01 \pm 0.04$  & $0.90 \pm 0.07$ & $-8.49 \pm 0.13$ & $-8.78 \pm 0.15$  \\
1.8 - 2.0 & $1.05 \pm 0.05$  & $0.90 \pm 0.20$ & $-8.86 \pm 0.15$ & $-9.00 \pm 0.19$  \\
2.0 - 2.2 & $0.9 \pm 0.09$   & $1.18 \pm 0.38$ & $-7.42 \pm 0.17$ & $-11.55 \pm 0.31$ \\
2.2 - 2.4 & $1.03 \pm 0.09$  & $0.59 \pm 0.26$ & $-8.62 \pm 0.26$ & $-5.92 \pm 0.39$  \\ \hline
\end{tabular}%
}
\end{table}

These results reinforce the distinction between G1 and G2 as bulge-dominated galaxies at different stages of evolution. G1 systems appear to be long-lived, star-forming systems with SFR–mass scaling similar to disk galaxies. In contrast, G2 galaxies exhibit properties consistent with recently quenched or quenching systems, gradually shifting from the BC to the RS. This pattern aligns with the evolutionary framework proposed in prior studies suggesting that galaxies transition through the green valley en route to quiescence \citep[e.g.,][]{2022MNRAS.509..567S}.

Furthermore, the persistence of G1 in the BC suggests that these galaxies may follow a distinct evolutionary track—possibly formed in situ with bulge-dominated morphology and sustained star formation. Meanwhile, G2 systems, which increase in mass and evolve toward quiescence over time, likely arise from major merger events, consistent with the observed build-up of massive ellipticals in the local Universe \citep[e.g.,][]{2007IAUS..235..381C, 2013Natur.498..338F, 2023MNRAS.519.2119Q}. The evolving SFMS positions of G1 and G2 thus provide further evidence that these two bulge-dominated families represent structurally and physically distinct populations, shaped by different evolutionary mechanisms.

\subsection{Comparing structural parameters}

Changes in star formation activity are often accompanied by structural transformations. To investigate this connection for the two bulge-dominated subpopulations identified earlier, we examine how their effective radius ($R_{\rm e}$) and Sérsic index ($n_{\rm s}$) evolve with redshift. Given that both G1 and G2 are morphologically classified as bulge-dominated galaxies, Sérsic profile parameters are a natural choice for quantifying their structure.

\begin{figure*}
    \centering
    \includegraphics[width=\textwidth]{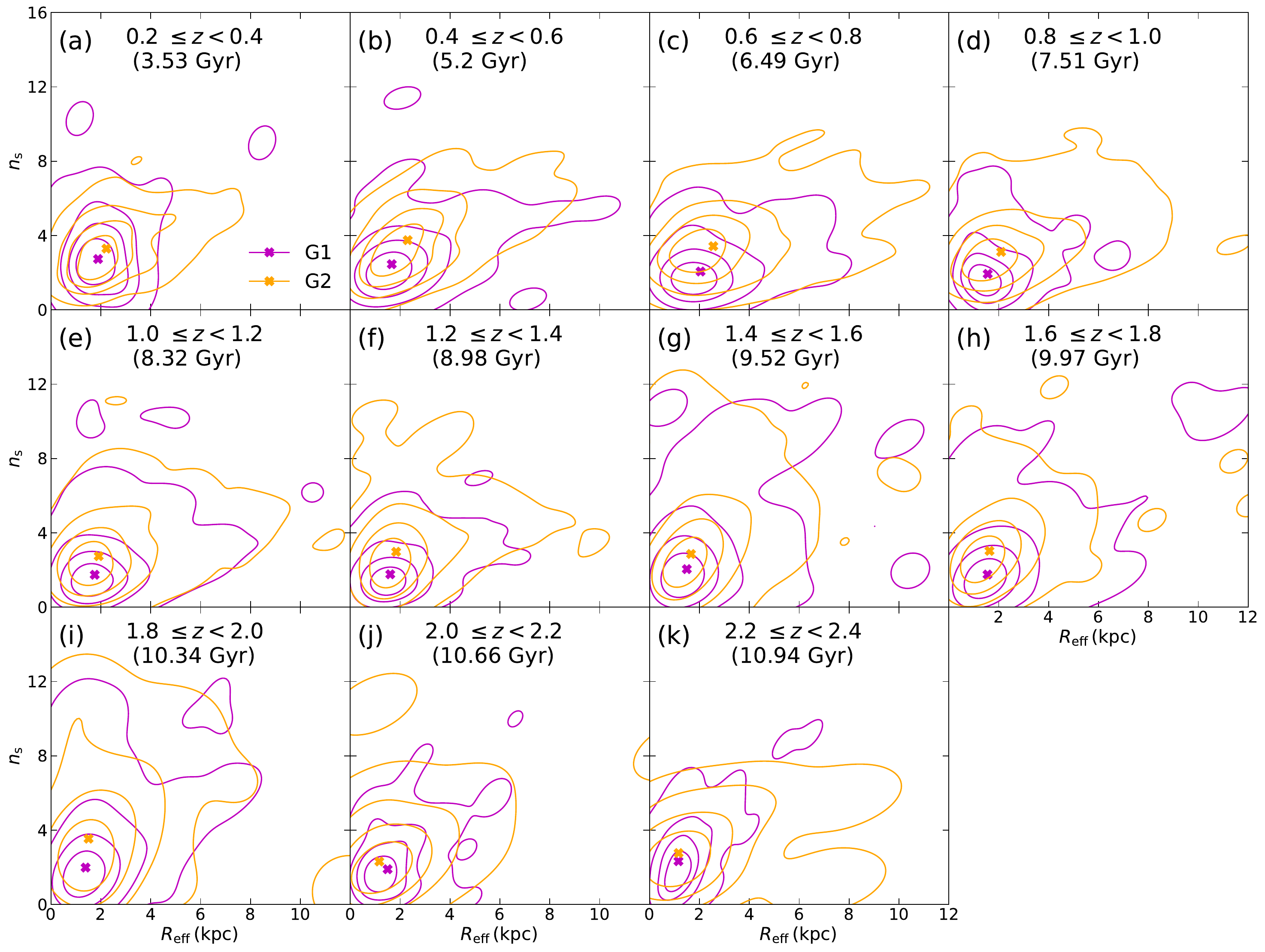}
    \caption{Contour lines of the G1 (magenta) and G2 (orange) distributions in the $n_{\rm s}$ \emph{vs.} $R_{\rm e}$ diagram, with each panel representing a different redshift bin.}
    \label{fig:bimodality_sersic}
\end{figure*}

In Fig.~\ref{fig:bimodality_sersic}, we show the $n_{\rm s}$ vs. $R_{\rm e}$ distributions of G1 (magenta) and G2 (orange) across redshift bins. The contour lines represent the 25th, 50th, 75th, and 90th percentiles of each distribution, and the colored crosses denote the mean values.

We find that G1 and G2 exhibit similar $R_{\rm e}$ values across redshifts, with an average difference of only $\Delta R_{\rm e}\,[{\rm G2 - G1}] = 0.2 \pm 0.1$\,kpc. However, the Sérsic index differs significantly: G2 galaxies are consistently more centrally concentrated than G1 systems, with an average offset of $\Delta n_{\rm s}\,[{\rm G2 - G1}] = 0.99 \pm 0.35$. This suggests that G2 galaxies have steeper central light profiles, which is often interpreted as a signature of dissipative processes such as gas-rich major mergers \citep[e.g.,][]{2009ApJS..181..135H, 2015MNRAS.447.3291C}.

Our findings support a scenario in which G2 galaxies result from merger-driven mass assembly. The higher Sérsic index values in G2—despite comparable effective radii—indicate that stellar mass is more centrally concentrated. This is consistent with simulations and observations showing that major mergers not only build stellar mass but also increase the central concentration of light, producing high-$n$ systems \citep{2018MNRAS.478.3994C, 2023ApJ...951..115E}.

In contrast, G1 systems exhibit lower Sérsic indices, indicative of more diffuse light profiles. Combined with their persistent star formation (as shown in Fig.~\ref{fig:bimodality_sfms}), this suggests a different formation path—possibly one tied to in-situ processes and angular momentum retention during collapse \citep{1969ApJ...155..393P, 2015ApJ...812...29T}. This dichotomy in structural properties reinforces the picture that G1 and G2 represent bulge-dominated systems at different evolutionary stages and shaped by distinct physical mechanisms. Additionally, our results are consistent with recent findings that galaxies in the green valley tend to have higher Sérsic indices, reflecting more concentrated stellar distributions as they transition toward quiescence \citep{2017ApJ...840...47B, 2019MNRAS.484.3022N}.

\section{Summary and Discussion}

In this work, we investigated the evolution of disk and bulge-dominated galaxies using a mass-complete sample from the CANDELS survey, limited to $H_{\rm mag} \leq 24$, $M_{\rm stellar} \geq 10^{9}\,{\rm M}_\odot$, and within the redshift range $0.2 \leq z \leq 2.4$. Adopting an unbiased, non-parametric morphological classification, we traced the structural and star formation properties of $\sim$14,000 galaxies over more than 10 Gyr of cosmic time. Our main goal was to dissect how morphology, star formation, and stellar mass co-evolve, with special attention to the emergence of different sub-populations among bulge-dominated galaxies. Our main findings are summarized as follows:
\begin{itemize}
    \item \textbf{sSFR divergence and bimodality}: At high redshift ($z \sim 2.4$), disk and bulge-dominated galaxies exhibit similar specific star formation rate (sSFR) distributions. However, with decreasing redshift, bulge-dominated galaxies evolve differently—developing a bimodal sSFR distribution below $z < 1.6$, while disk galaxies retain a unimodal, Gaussian-like distribution with a tail toward low sSFR.

    \item \textbf{Co-evolution of SFR and stellar mass}: While SFR distributions of disks and bulge-dominated galaxies remain similar across redshifts, the divergence in sSFR is primarily driven by differences in stellar mass. At $z \sim 0.2$, the SFR gap between the two is only $\sim$0.3 dex—far smaller than the $\sim$1 dex gap observed in the local Universe—suggesting a recent, rapid decline in star formation of bulge-dominated galacies over the last $\sim$1.5 Gyr.

    \item \textbf{Mass assembly through mergers}: The stellar mass distributions of bulge-dominated systems evolve from being disk-like at high redshift to strongly bimodal at $z < 1.6$, with a new massive population emerging. Simultaneously, disks show decreasing average mass despite steady SFRs, suggesting that massive disk galaxies are being destroyed—likely by major mergers—and being transformed into massive bulge-dominated systems.

    \item \textbf{Morphological transition is mass-dependent}: The evolution of disk and bulge-dominated fractions varies significantly with stellar mass. While low-mass galaxies tend to retain disk morphologies toward lower redshifts, high-mass galaxies show a declining disk fraction and rising bulge-dominated fraction, consistent with a merger-driven transformation pathway that preferentially affects massive systems.

    \item \textbf{Two bulge-dominated evolutionary tracks}: We identify and characterize two structural families of bulge-dominated galaxies. G1 systems are star-forming, lower-mass bulge-dominated galaxies with properties similar to disks, occupying the blue cloud at all redshifts. G2 systems are more massive, less star-forming, and increasingly dominate at $z < 1$. Their higher Sérsic indices (by $\Delta n_{\rm s} \sim 1$) and similar effective radii suggest more concentrated light profiles, supporting a formation pathway involving major mergers followed by rapid quenching.

    \item \textbf{Quenching sequence in the SFMS}: The location of G1 and G2 in the star formation main sequence (SFMS) reinforces their interpretation as distinct evolutionary stages. G1 galaxies populate the blue cloud irrespective of redshift, while G2 galaxies transition into the green valley and red sequence over time—mirroring a quenching sequence and consistent with trends observed in the local Universe.
\end{itemize}

Despite the strengths of this analysis, several caveats must be acknowledged. Our conclusions rely on sSFR and SFR estimates from SED fitting, which—though robust for CANDELS—still carry uncertainties, especially for dusty or low-SFR systems. Additionally, the use of fixed, local-Universe criteria to define the SFMS regions (blue cloud, green valley, red sequence) introduces potential bias at higher redshift. The interpretation of G1 and G2 as distinct evolutionary stages is statistically supported, but disentangling causality (e.g., mergers vs. fading) requires kinematic data and high-resolution simulations. Future work using JWST data, integral field spectroscopy, and cosmological simulations will allow more direct tests of these mechanisms and tighter constraints on structural evolution at high redshift.

\section*{Acknowledgements}
The authors acknowledge the comments of the anonymous referee, that significantly helped improving this work, especially regarding on how it is presented. VMS and RRdC acknowledge the support from FAPESP through the grants 2020/16243-3 and 2020/15245-2. VMS also acknowledge financial support from ANID - MILENIO - NCN2024\_112IF acknowledges support from the Spanish Ministry of Science, Innovation and Universities (MCIU), through grant PID2019-104788GB-I00. 

This work utilizes observations from the CANDELS survey, obtained through the Mikulski Archive for Space Telescopes (MAST). CANDELS is a legacy program (GO 12060) led by the University of California, Santa Cruz, and partially supported by the Hubble Space Telescope (HST) under NASA contract NAS 5-26555. We acknowledge the use of HST data and thank the CANDELS team for making these high-quality data publicly available.

We also acknowledge the use of several Python libraries that were instrumental in the data analysis and visualization for this study. Specifically, we utilized NumPy \citep{numpy2020}, SciPy \citep{scipy2020}, pandas \citep{pandas2020}, Matplotlib \citep{matplotlib2007}, and scikit-learn \citep{scikit-learn2011}. These open-source tools provided robust and efficient functionalities for numerical computations, statistical modeling, data manipulation, and visualization, enabling reproducible and rigorous analysis.
\section*{Data Availability}

The data underlying this paper were accessed from CANDELS database. The data underlying this article will be shared on request to the corresponding author.



\bibliographystyle{mnras}
\bibliography{example} 




\appendix

\section{Comparison between different star formation rate estimates}
\label{sec:appendix_mstarsfr}

\begin{figure}
    \centering
    \includegraphics[width=0.8\columnwidth]{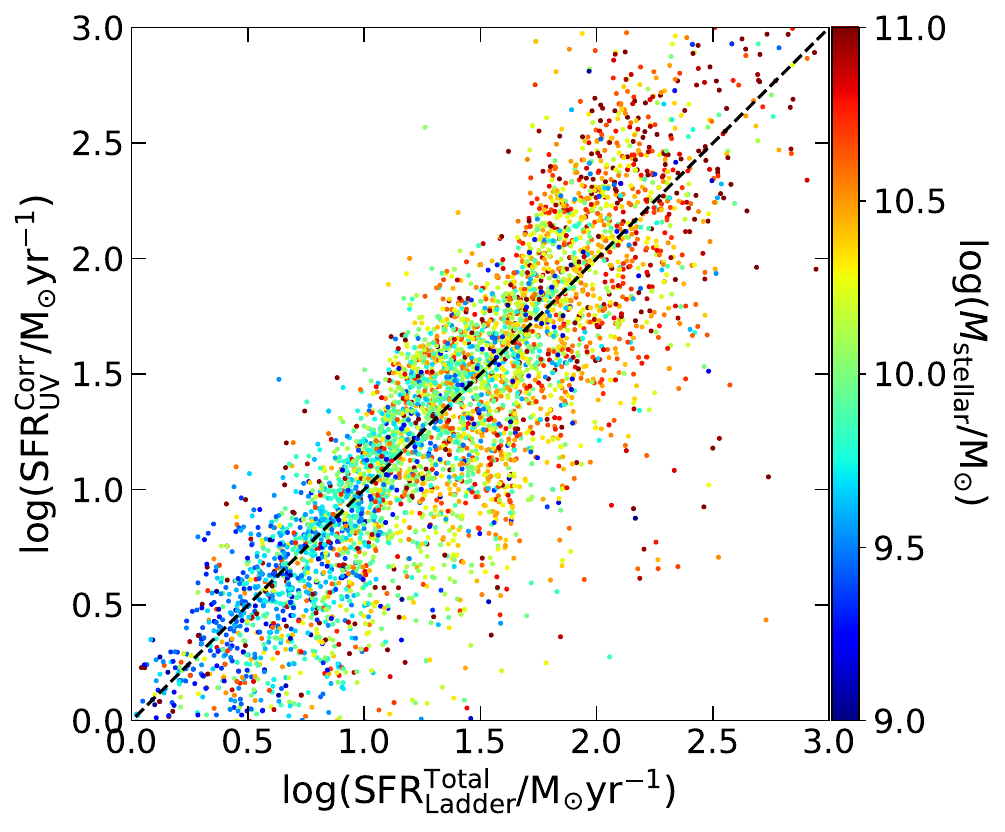}
    \caption{Relation between ${\rm SFR}_{\rm UV}^{\rm Corr}$ and ${\rm SFR}_{\rm Ladder}^{\rm Total}$. Galaxies are colored according to their stellar mass. The dashed black line shows the equality line.}
    \label{fig:SFR_comparison}
\end{figure}

In this appendix, we explore the agreement between the SFR estimates obtained through the combination of UV+IR (${\rm SFR}_{\rm Ladder}^{\rm Total}$) observations with those estimated only using UV, and then applying a dust correction (${\rm SFR}_{\rm UV}^{\rm Corr}$. For this comparison we selected all galaxies from the CANDELS matching our sample selection criteria, i.e. $Hmag \leq 24$ and $M_{\rm stellar} \geq 10^{9}{\rm M}_{\odot}$. Moreover, we select only galaxies with SFR-``ladder\_type'' equal to 1 or 2 in \cite{barro2019candels} catalog, corresponding to systems for which the SFR is derived with both UV and IR observations, although using different approaches for 1 and 2. This results in a sample of 5,143 galaxies. We show in Fig.~\ref{fig:SFR_comparison} the relation between ${\rm SFR}_{\rm UV}^{\rm Corr}$ and ${\rm SFR}_{\rm Ladder}^{\rm Total}$, with galaxies colored according to their stellar mass, which is retrieved from \cite{santini2015stellar}.

Exploring Fig.~\ref{fig:SFR_comparison}, we note, on average, an agreement between the two estimates. Despite a large scatter of $\sim 0.3$, this is comparable to the expected uncertainty in the estimate itself. We therefore decide to use ${\rm SFR}_{\rm UV}^{\rm Corr}$ to enhance the number of galaxies in our sample, without loosing robustness. To ensure that our conclusions do not depend on the SFR estimate used, we consider only differences greater than 0.3 dex.

\section{Quantifying properties of distributions}

In this appendix, we present relevant tables describing statistical properties of the distributions shown throughout this paper. We decide to avoid including them in the main text to keep a better reading flow. Tables~\ref{tab:ssfr_description}, \ref{tab:sfr_description}, and \ref{tab:mstellar_description} present the mean, median, mode, $\rm 1^{st}$ ($Q_{25\%})$ and $3^{\rm rd}$ ($Q_{75\%}$) quartiles, scatter calculated using the quartiles ($Q_{\sigma}$), kurtosis excess and skewness of the disk and bulge-dominated galaxies distribution presented in Figs.~\ref{fig:s_d_ssfr}, \ref{fig:s_d_sfr}, \ref{fig:s_d_mstellar}, respectively. In Table \ref{tab:bimodality_description}, we detail the mean, standard deviation and weight of each Gaussian component, found by applying a Bayesian Gaussian Mixture Model to the sSFR distribution of bulge-dominated galaxies (red curves in Fig.~\ref{fig:s_d_ssfr}) as a function of redshift. We highlight in red the three highest redshift bins as the bimodality in the sSFR distribution of bulge-dominated galaxies only become prominent below $z < 1.6$, such that the highlighted results should be addressed with caution. By definition, $\rm G1$ and $\rm G2$ comprises, respectively, bulge-dominated galaxies with sSFR comparable and significant smaller than disks. The G1 and G2 components are shown as purple and orange dashed lines in Fig.~\ref{fig:bimodality_sfms}.  

\begin{table}
\caption{Properties of the sSFR distributions shown in Fig.~\ref{fig:s_d_ssfr}. Top and bottom half show the results for disk and bulge-dominated galaxies, respectively. The content of the columns are as follows: (1) redshift bin considered; (2) distribution mean; (2) second quartile (i.e. median); (3) Mode (i.e. the value with most occurrences; (4) first quartile; (5) third quartile; (6) standard deviation calculated using quartiles; (7) the excess in kurtosis (i.e. 0 means Gaussian); and (8) the distribution skewness.}
\label{tab:ssfr_description}

\resizebox{\columnwidth}{!}{%
\begin{tabular}{ccccccccc}

\hline
\multicolumn{9}{c}{sSFR distribution of disk galaxies}                                                                                                                                                 \\ \hline
\multicolumn{1}{c|}{z bin}     & Mean   & Median & Mode   & $Q_{25\%}$ & $Q_{75\%}$ & $Q_{\sigma}$ & \begin{tabular}[c]{@{}c@{}}Kurtosis\\ Excess\end{tabular} & Skewness \\ \hline
\multicolumn{1}{c|}{0.2 - 0.4} & -9.43  & -9.37  & -9.29  & -9.81      & -9.05      & 0.56         & 2.05                                                      & -0.14    \\
\multicolumn{1}{c|}{0.4 - 0.6} & -9.32  & -9.19  & -8.96  & -9.63      & -8.89      & 0.55         & 2.76                                                      & -0.90    \\
\multicolumn{1}{c|}{0.6 - 0.8} & -9.23  & -9.08  & -8.84  & -9.49      & -8.81      & 0.50         & 4.00                                                      & -1.42    \\
\multicolumn{1}{c|}{0.8 - 1.0} & -9.17  & -9.00  & -8.77  & -9.41      & -8.76      & 0.48         & 4.25                                                      & -1.50    \\
\multicolumn{1}{c|}{1.0 - 1.2} & -9.11  & -8.92  & -8.95  & -9.38      & -8.67      & 0.52         & 4.03                                                      & -1.62    \\
\multicolumn{1}{c|}{1.2 - 1.4} & -9.00  & -8.79  & -8.72  & -9.23      & -8.58      & 0.48         & 4.63                                                      & -1.71    \\
\multicolumn{1}{c|}{1.4 - 1.6} & -9.15  & -8.81  & -8.57  & -9.56      & -8.51      & 0.78         & 2.05                                                      & -1.42    \\
\multicolumn{1}{c|}{1.6 - 1.8} & -8.79  & -8.57  & -8.44  & -8.90      & -8.39      & 0.37         & 4.83                                                      & -1.78    \\
\multicolumn{1}{c|}{1.8 - 2.0} & -8.86  & -8.57  & -8.57  & -9.04      & -8.35      & 0.50         & 3.46                                                      & -1.79    \\
\multicolumn{1}{c|}{2.0 - 2.2} & -8.76  & -8.50  & -8.39  & -8.80      & -8.27      & 0.38         & 4.74                                                      & -2.04    \\
\multicolumn{1}{c|}{2.2 - 2.4} & -8.86  & -8.51  & -8.60  & -8.88      & -8.29      & 0.43         & 3.75                                                      & -1.93    \\ \hline
\multicolumn{9}{c}{sSFR distribution of bulge-dominated galaxies}                                                                                                                                             \\ \hline
\multicolumn{1}{c|}{z bin}     & Mean   & Median & Mode   & $Q_{25\%}$ & $Q_{75\%}$ & $Q_{\sigma}$ & \begin{tabular}[c]{@{}c@{}}Kurtosis\\ Excess\end{tabular} & Skewness \\
\multicolumn{1}{c|}{0.2 - 0.4} & -10.32 & -10.49 & -10.60 & -10.93     & -9.81      & 0.83         & -0.44                                                     & 0.60     \\
\multicolumn{1}{c|}{0.4 - 0.6} & -10.02 & -10.09 & -10.69 & -10.69     & -9.41      & 0.95         & 0.38                                                      & 0.21     \\
\multicolumn{1}{c|}{0.6 - 0.8} & -10.00 & -10.12 & -10.41 & -10.66     & -9.24      & 1.05         & 0.11                                                      & 0.11     \\
\multicolumn{1}{c|}{0.8 - 1.0} & -9.72  & -9.75  & -10.33 & -10.44     & -8.90      & 1.14         & -0.42                                                     & -0.26    \\
\multicolumn{1}{c|}{1.0 - 1.2} & -9.49  & -9.19  & -8.74  & -10.25     & -8.74      & 1.12         & 0.00                                                      & -0.71    \\
\multicolumn{1}{c|}{1.2 - 1.4} & -9.26  & -8.93  & -8.68  & -9.98      & -8.59      & 1.03         & 0.85                                                      & -0.90    \\
\multicolumn{1}{c|}{1.4 - 1.6} & -9.33  & -8.85  & -8.65  & -10.00     & -8.50      & 1.11         & 0.43                                                      & -1.09    \\
\multicolumn{1}{c|}{1.6 - 1.8} & -8.90  & -8.60  & -8.34  & -9.64      & -8.33      & 0.97         & -0.12                                                     & -0.73    \\
\multicolumn{1}{c|}{1.8 - 2.0} & -8.84  & -8.46  & -8.46  & -9.31      & -8.26      & 0.78         & 0.87                                                      & -1.22    \\
\multicolumn{1}{c|}{2.0 - 2.2} & -8.64  & -8.40  & -8.39  & -8.71      & -8.24      & 0.35         & 4.84                                                      & -1.95    \\
\multicolumn{1}{c|}{2.2 - 2.4} & -8.60  & -8.44  & -8.52  & -8.68      & -8.22      & 0.33         & 3.63                                                      & -1.79    \\ \hline
\end{tabular}%
}
\end{table}

\begin{table}
\centering
\caption{Same as Table~\ref{tab:ssfr_description}, but for the SFR distributions.}
\label{tab:sfr_description}

\resizebox{\columnwidth}{!}{%
\begin{tabular}{ccccccccc}
\hline
\multicolumn{9}{c}{SFR distribution of disk galaxies}                                                                                                                                       \\ \hline
\multicolumn{1}{c|}{z bin}     & Mean  & Median & Mode  & $Q_{25\%}$ & $Q_{75\%}$ & $Q_{\sigma}$ & \begin{tabular}[c]{@{}c@{}}Kurtosis\\ Excess\end{tabular} & Skewness \\ \hline
\multicolumn{1}{c|}{0.2 - 0.4} & -0.03 & -0.02  & 0.16  & -0.45      & 0.39       & 0.63         & 0.63                                                      & 0.07     \\
\multicolumn{1}{c|}{0.4 - 0.6} & 0.18  & 0.25   & 0.36  & -0.22      & 0.68       & 0.68         & 1.48                                                      & -0.71    \\
\multicolumn{1}{c|}{0.6 - 0.8} & 0.29  & 0.33   & 0.41  & -0.04      & 0.70       & 0.55         & 2.14                                                      & -0.74    \\
\multicolumn{1}{c|}{0.8 - 1.0} & 0.41  & 0.44   & 0.22  & 0.12       & 0.83       & 0.53         & 2.96                                                      & -0.94    \\
\multicolumn{1}{c|}{1.0 - 1.2} & 0.54  & 0.58   & 0.58  & 0.24       & 1.02       & 0.57         & 2.51                                                      & -0.95    \\
\multicolumn{1}{c|}{1.2 - 1.4} & 0.66  & 0.71   & 0.65  & 0.45       & 0.97       & 0.38         & 4.00                                                      & -1.24    \\
\multicolumn{1}{c|}{1.4 - 1.6} & 0.64  & 0.90   & 0.97  & 0.43       & 1.10       & 0.49         & 2.47                                                      & -1.35    \\
\multicolumn{1}{c|}{1.6 - 1.8} & 1.04  & 1.08   & 1.13  & 0.93       & 1.22       & 0.21         & 5.48                                                      & -1.32    \\
\multicolumn{1}{c|}{1.8 - 2.0} & 0.96  & 1.07   & 1.12  & 0.92       & 1.27       & 0.26         & 2.80                                                      & -1.29    \\
\multicolumn{1}{c|}{2.0 - 2.2} & 1.11  & 1.15   & 1.09  & 1.00       & 1.42       & 0.31         & 4.68                                                      & -1.55    \\
\multicolumn{1}{c|}{2.2 - 2.4} & 1.00  & 1.21   & 1.24  & 1.00       & 1.39       & 0.29         & 3.50                                                      & -1.58    \\ \hline
\multicolumn{9}{c}{SFR distribution of bulge-dominated galaxies}                                                                                                                                           \\ \hline
\multicolumn{1}{c|}{z bin}     & Mean  & Median & Mode  & $Q_{25\%}$ & $Q_{75\%}$ & $Q_{\sigma}$ & \begin{tabular}[c]{@{}c@{}}Kurtosis\\ Excess\end{tabular} & Skewness \\
\multicolumn{1}{c|}{0.2 - 0.4} & -0.31 & -0.35  & -0.47 & -0.81      & -0.35      & 0.72         & -0.09                                                     & 0.23     \\
\multicolumn{1}{c|}{0.4 - 0.6} & 0.00  & -0.01  & 0.24  & -0.43      & -0.01      & 0.64         & 2.39                                                      & -0.05    \\
\multicolumn{1}{c|}{0.6 - 0.8} & 0.17  & 0.12   & 0.12  & -0.30      & 0.12       & 0.70         & 0.97                                                      & -0.02    \\
\multicolumn{1}{c|}{0.8 - 1.0} & 0.27  & 0.26   & 0.24  & -0.13      & 0.26       & 0.60         & 2.32                                                      & -0.41    \\
\multicolumn{1}{c|}{1.0 - 1.2} & 0.41  & 0.47   & 0.57  & 0.10       & 0.47       & 0.52         & 2.71                                                      & -0.92    \\
\multicolumn{1}{c|}{1.2 - 1.4} & 0.63  & 0.67   & 0.68  & 0.37       & 0.67       & 0.39         & 5.39                                                      & -1.03    \\
\multicolumn{1}{c|}{1.4 - 1.6} & 0.61  & 0.88   & 0.97  & 0.33       & 0.88       & 0.59         & 1.76                                                      & -1.25    \\
\multicolumn{1}{c|}{1.6 - 1.8} & 1.08  & 1.04   & 1.07  & 0.88       & 1.04       & 0.26         & 3.44                                                      & -0.35    \\
\multicolumn{1}{c|}{1.8 - 2.0} & 1.09  & 1.11   & 1.06  & 0.95       & 1.11       & 0.30         & 2.16                                                      & -0.61    \\
\multicolumn{1}{c|}{2.0 - 2.2} & 1.26  & 1.19   & 1.16  & 1.08       & 1.19       & 0.35         & 4.78                                                      & -1.21    \\
\multicolumn{1}{c|}{2.2 - 2.4} & 1.26  & 1.27   & 1.23  & 1.09       & 1.27       & 0.31         & 1.47                                                      & -0.54    \\ \hline
\end{tabular}%
}
\end{table}

\begin{table}
\centering
\caption{Same as Table~\ref{tab:ssfr_description}, but for the $M_{\rm stellar}$ distributions.}
\label{tab:mstellar_description}

\resizebox{\columnwidth}{!}{%
\begin{tabular}{ccccccccc}
\hline
\multicolumn{9}{c}{$M_{\rm stellar}$ distribution of disk galaxies}                                                                                                     \\ \hline
\multicolumn{1}{c|}{z bin}     & Mean  & Median & Mode  & $Q_{25\%}$ & $Q_{75\%}$ & $Q_{\sigma}$ & \begin{tabular}[c]{@{}c@{}}Kurtosis\\ Excess\end{tabular} & Skewness \\ \hline
\multicolumn{1}{c|}{0.2 - 0.4} & 9.41  & 9.34   & 9.12  & 9.13       & 9.61       & 0.35         & 0.98                                                      & 1.14     \\
\multicolumn{1}{c|}{0.4 - 0.6} & 9.52  & 9.41   & 9.15  & 9.17       & 9.80       & 0.46         & 0.48                                                      & 0.92     \\
\multicolumn{1}{c|}{0.6 - 0.8} & 9.52  & 9.42   & 9.12  & 9.18       & 9.77       & 0.43         & 0.06                                                      & 0.87     \\
\multicolumn{1}{c|}{0.8 - 1.0} & 9.59  & 9.50   & 9.14  & 9.21       & 9.90       & 0.51         & -0.35                                                     & 0.71     \\
\multicolumn{1}{c|}{1.0 - 1.2} & 9.66  & 9.57   & 9.17  & 9.25       & 9.97       & 0.54         & -0.36                                                     & 0.65     \\
\multicolumn{1}{c|}{1.2 - 1.4} & 9.67  & 9.58   & 9.33  & 9.30       & 9.98       & 0.50         & -0.34                                                     & 0.66     \\
\multicolumn{1}{c|}{1.4 - 1.6} & 9.81  & 9.74   & 9.55  & 9.43       & 10.09      & 0.49         & -0.05                                                     & 0.73     \\
\multicolumn{1}{c|}{1.6 - 1.8} & 9.84  & 9.72   & 9.41  & 9.47       & 10.14      & 0.49         & 0.01                                                      & 0.83     \\
\multicolumn{1}{c|}{1.8 - 2.0} & 9.83  & 9.75   & 9.61  & 9.51       & 10.09      & 0.43         & 0.46                                                      & 0.84     \\
\multicolumn{1}{c|}{2.0 - 2.2} & 9.88  & 9.80   & 9.64  & 9.53       & 10.12      & 0.44         & 0.45                                                      & 0.84     \\
\multicolumn{1}{c|}{2.2 - 2.4} & 9.87  & 9.81   & 9.73  & 9.62       & 10.08      & 0.34         & 0.59                                                      & 0.72     \\ \hline
\multicolumn{9}{c}{$M_{\rm stellar}$ distribution of bulge-dominated galaxies}                                                                                               \\ \hline
\multicolumn{1}{c|}{z bin}     & Mean  & Median & Mode  & $Q_{25\%}$ & $Q_{75\%}$ & $Q_{\sigma}$ & \begin{tabular}[c]{@{}c@{}}Kurtosis\\ Excess\end{tabular} & Skewness \\
\multicolumn{1}{c|}{0.2 - 0.4} & 10.00 & 10.02  & 10.23 & 9.64       & 10.34      & 0.52         & -0.58                                                     & 0.08     \\
\multicolumn{1}{c|}{0.4 - 0.6} & 10.03 & 10.07  & 10.02 & 9.56       & 10.52      & 0.71         & -1.08                                                     & -0.10    \\
\multicolumn{1}{c|}{0.6 - 0.8} & 10.18 & 10.24  & 10.27 & 9.77       & 10.60      & 0.61         & -0.74                                                     & -0.28    \\
\multicolumn{1}{c|}{0.8 - 1.0} & 9.99  & 10.01  & 9.31  & 9.41       & 10.48      & 0.79         & -1.07                                                     & 0.15     \\
\multicolumn{1}{c|}{1.0 - 1.2} & 9.90  & 9.78   & 9.43  & 9.35       & 10.45      & 0.82         & -1.14                                                     & 0.36     \\
\multicolumn{1}{c|}{1.2 - 1.4} & 9.89  & 9.78   & 9.35  & 9.33       & 10.45      & 0.83         & -1.16                                                     & 0.35     \\
\multicolumn{1}{c|}{1.4 - 1.6} & 9.94  & 9.82   & 9.69  & 9.46       & 10.46      & 0.74         & -0.98                                                     & 0.41     \\
\multicolumn{1}{c|}{1.6 - 1.8} & 9.99  & 9.78   & 9.52  & 9.47       & 10.57      & 0.81         & -1.07                                                     & 0.49     \\
\multicolumn{1}{c|}{1.8 - 2.0} & 9.93  & 9.74   & 9.64  & 9.49       & 10.41      & 0.68         & -0.53                                                     & 0.68     \\
\multicolumn{1}{c|}{2.0 - 2.2} & 9.90  & 9.73   & 9.66  & 9.52       & 10.19      & 0.49         & 0.10                                                      & 0.91     \\
\multicolumn{1}{c|}{2.2 - 2.4} & 9.87  & 9.80   & 9.68  & 9.57       & 10.07      & 0.37         & 1.09                                                      & 0.95     \\ \hline
\end{tabular}%
}
\end{table}

\begin{table}
\caption{Description of the two Gaussian components describing the sSFR distribution of bulge-dominated galaxies. Column (1) shows the redshift bin considered; columns (2) and (3) shows the mean for G1 and G2, respectively; columns (4) and (5) do the same, but for the standard deviation, also respectively for G1 and G2; and columns (6) and (7) presents the weights of each component, for G1 and G2, in that order. Uncertainties in the mean and in the standard deviation are calculated using a bootstrap technique with 1000 repetitions. We highlight in red the redshift bins corresponding to panels (i), (j) and (k), where the bimodality is not as prominent as in the other panels, thus the results should be taken with extra caution.}
\label{tab:bimodality_description}
\resizebox{\columnwidth}{!}{%
\begin{tabular}{c|cc|cc|cc}
\hline
          & \multicolumn{2}{c|}{$\mu_{i}$} & \multicolumn{2}{c|}{$\sigma_{i}$} & \multicolumn{2}{c}{$w_{i}$} \\ \hline
z bin     & G1            & G2             & G1              & G2              & G1           & G2           \\ \hline
0.2 - 0.4 & $-9.56\pm0.06$ & $-10.84\pm0.03$ & $0.47\pm0.04$    & $0.31\pm0.02$    & $0.44$       & $0.56$       \\
0.4 - 0.6 & $-9.38\pm0.19$ & $-10.72\pm0.12$ & $0.48\pm0.05$    & $0.4\pm0.09$     & $0.45$       & $0.55$       \\
0.6 - 0.8 & $-9.23\pm0.26$ & $-10.66\pm0.14$ & $0.45\pm0.06$    & $0.46\pm0.11$    & $0.31$       & $0.69$       \\
0.8 - 1.0 & $-9.13\pm0.29$ & $-10.59\pm0.17$ & $0.42\pm0.08$    & $0.49\pm0.11$    & $0.40$       & $0.60$       \\
1.0 - 1.2 & $-9.05\pm0.3$  & $-10.51\pm0.22$ & $0.39\pm0.09$    & $0.55\pm0.16$    & $0.42$       & $0.58$       \\
1.2 - 1.4 & $-8.98\pm0.32$ & $-10.41\pm0.29$ & $0.35\pm0.11$    & $0.61\pm0.18$    & $0.43$       & $0.57$       \\
1.4 - 1.6 & $-8.91\pm0.34$ & $-10.41\pm0.28$ & $0.34\pm0.11$    & $0.65\pm0.21$    & $0.52$       & $0.48$       \\
1.6 - 1.8 & $-8.85\pm0.36$ & $-10.35\pm0.32$ & $0.34\pm0.12$    & $0.62\pm0.23$    & $0.63$       & $0.37$       \\
\rowcolor[HTML]{FD6864} 
1.8 - 2.0 & $-8.79\pm0.38$ & $-10.31\pm0.32$ & $0.33\pm0.12$    & $0.66\pm0.22$    & $0.64$       & $0.36$       \\
\rowcolor[HTML]{FD6864} 
2.0 - 2.2 & $-8.75\pm0.38$ & $-10.25\pm0.35$ & $0.33\pm0.09$    & $0.71\pm0.25$    & $0.73$       & $0.27$       \\
\rowcolor[HTML]{FD6864} 
2.2 - 2.4 & $-8.71\pm0.38$ & $-10.23\pm0.35$ & $0.32\pm0.09$    & $0.69\pm0.24$    & $0.82$       & $0.18$       \\ \hline
\end{tabular}%
}
\end{table}

\section{Defining the number of components in the Gaussian Mixture Model}
\label{sec_app:GMM}
\begin{figure}
    \centering
    \includegraphics[width=0.8\columnwidth]{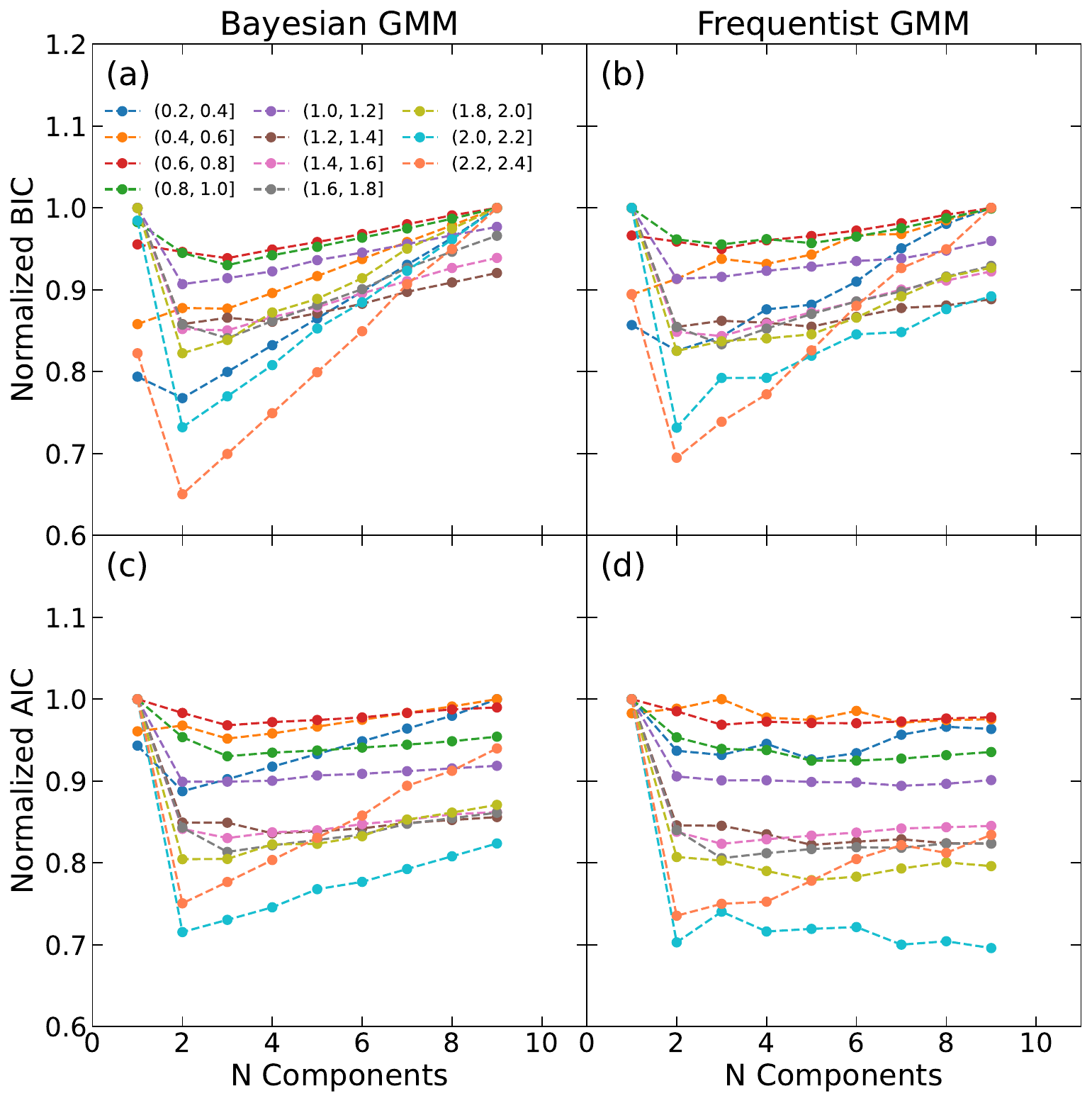}
    \caption{The normalized BIC (top panels) and AIC (bottom panels) values as a function of the number of components used in the GMM to fit the bulge-dominated galaxies sSFR distributions shown in Fig.~\ref{fig:s_d_ssfr}. Each colored line represents a given redshift bin. Left panels show the results when using the Bayesian GMM, whereas the right-hand panels do the same for the frequentist GMM. Notably, irrespective of criteria or using bayesian/frequentist GMM, we find that most of the distributions are well described by two components.} 
    \label{fig:BIC_AIC}
\end{figure}

In Fig.~\ref{fig:BIC_AIC}, we show the variation of BIC (top panels) and AIC (bottom panels) as a function of the number of components used in the GMM fitting. Each colored line represents the results for a given redshift bin. Left panels show the results when using the Bayesian approach, whereas the right-hand side do the same for the frequentist GMM. Since the number of bulge-dominated galaxies varies with redshift, we normalize both BIC and AIC by the maximum value found when varying the number of components from 1 to 10. We find that, irrespective of redshift bin considered, the approach (Bayesian/Frequentist), and even the criteria (AIC/BIC) adopted, most cases are well described by a bimodal (n=2) distribution. Even in the case where the minimum AIC/BIC happens at $n \neq 2$, the difference for $n=2$ is significantly small. Therefore, we decide to adopt $n = 2$ for the GMM, irrespective of redshift. 

\bsp	
\label{lastpage}
\end{document}